\documentclass[a4paper,12pt]{article}
\usepackage{amsmath}
\usepackage{amssymb}
\usepackage{graphicx}
\usepackage{verbatim}
\usepackage{color}
\usepackage{subfigure}
\usepackage{array}
\usepackage{indentfirst}
\usepackage[numbers]{natbib}
\usepackage{chapterbib}
\usepackage[english]{babel}
\usepackage{fancyhdr}
\usepackage{multirow}
\usepackage[acronym,nomain]{glossaries}
\usepackage{xspace}
\usepackage{url}
\usepackage{eurosym}
\usepackage{lineno}
\usepackage{color}

\newcommand{\gco}{\,g~CO$_{2}\textrm{e}$\xspace}
\newcommand{\kco}{\,kg~CO$_{2}\textrm{e}$\xspace}
\newcommand{\tco}{\,t~CO$_{2}\textrm{e}$\xspace}
\newcommand{\ktco}{\,kt~CO$_{2}\textrm{e}$\xspace}
\newcommand{\tcoperyr}{\,t~CO$_{2}\textrm{e}$/yr\xspace}
\newcommand{\tcopermeuro}{\,t~CO$_{2}\textrm{e}$/M\euro\xspace}
\newcommand{\kcoperkeuro}{\,kg~CO$_{2}\textrm{e}$/k\euro\xspace}
\newcommand{\gcopergb}{\,g~CO$_{2}\textrm{e}$/GB\xspace}
\newcommand{\kgcoperkwh}{\,kg CO$_2$e/kWh\xspace}
\newcommand{\kgcopermsq}{\,kg CO$_2$e/m$^2$\xspace}
\newcommand{\kgcoperm}{\,kg~CO$_2$e/m$^{3}$\xspace}
\newcommand{\co}{CO$_2$\xspace}
\newcommand{\coe}{CO$_{2}\textrm{e}$\xspace}
\newcommand{\km}{\,km\xspace}
\newcommand{\degree}{\,$^\circ$C\xspace}
\newcommand{\gestot}{$3258$}
\newcommand{\gestotxl}{$7418$}

\newacronym{ghg}{GHG}{greenhouse gas}
\newacronym{bc}{BC}{Bilan Carbone}

\begin{document}

% Titre
\title{The carbon footprint of IRAP}
\author{Pierrick Martin, Sylvie Brau-Nogu\'e, Mickael Coriat, \\Philippe Garnier, Annie Hughes, J\"urgen Kn\"odlseder, Luigi Tibaldo \\IRAP, Universit\'e de Toulouse, CNRS, CNES, F-31028 Toulouse, France}
\date{25 April 2022}
\maketitle
\newpage

% Summary
\section{Summary}

\noindent We present an assessment of the greenhouse gases (GHG) emissions of the Institute for Research in Astrophysics and
Planetology (IRAP), located in Toulouse (France). It was performed following the established ``Bilan Carbone'' methodology, over a large scope compared to similar previous studies, 
including in particular the contribution from the purchase of goods and services as well as IRAP's use of external research infrastructures, such as ground-based observatories and space-borne facilities. The carbon footprint of the institute for the reference year 2019 is $7400 \pm
900$\tco. If we exclude the contribution from external research
infrastructures to focus on a restricted perimeter over which the
institute has some operational control, IRAP's emissions in 2019
amounted to $3300 \pm 400$\tco.\\

\noindent Over the restricted perimeter, the contribution from purchasing goods
and services is dominant, about 40\% of the total, slightly exceeding
the contribution from professional travel including hotel stays,
which accounts for 38\%. Local infrastructures make a smaller contribution to IRAP's carbon footprint, about 25\% over the
restricted perimeter. We note that this repartition may be
 specific to IRAP, since the energy used to produce the
 electricity and heating has a relatively low carbon footprint. Over the full perimeter, the large share from the use of ground-based observatories and space-borne facilities and the fact that the majority of IRAP purchases are related to instrument development indicate that research infrastructures represent the most significant challenge for reducing the carbon footprint of research at our institute.\\

\noindent With $\sim260$ staff members employed at our institute, our results imply
that performing research in astronomy and astrophysics at IRAP according to the
standards of 2019 produces average GHG emissions of 28\tcoperyr per person
involved in that activity.
This figure lies well above the target global average budget of
2\tcoperyr per capita by 2050. However, the footprint is spread across a variety of
social and economic sectors, and so are the benefits of the research activity.
As a consequence, the emission reduction to be
achieved by scientific research should be made on open and democratic grounds, in
a debate reaching beyond scientific communities and informed by facts and figures
such as those presented here.\\

\noindent Regardless of the exact reduction goals for our community, the magnitude of the challenge and the necessity to quickly 
engage into an effective transition calls for acting on all aspects of the
problem: lowering the carbon intensity of our activities, reducing their
pace, and shifting our work practices towards less
emission-intensive options. At the level of our own institute, we show that emissions in
the restricted perimeter can be reduced by up to 30\%, by changing our
traveling habits and adopting different practices for
commuting. Revising the criteria for the purchase of goods and
services could provide an additional significant reduction. At the
community level, the most urgent objective should be to lower the carbon footprint of research infrastructures and avoid that it keeps growing with the deployment of new facilities.\\

% TOC
\setcounter{tocdepth}{2}
\tableofcontents
\newpage

% Introduction
\section{Introduction}

\noindent In a 2018 Special Report `Global warming of
1.5\degree' \cite{ipcc:2018}, the Intergovernmental Panel on Climate
Change (IPCC) presented evidence that anthropogenic emissions, such as
greenhouse gases (GHG) and aerosols and their precursors, have
increased the global mean surface temperature by approximately
+1.0\degree\ (with a likely range of 0.8-1.2\degree) above
1850-1900 pre-industrial levels. The increase in carbon dioxide
content of the atmosphere is unprecedented over geological times. From
past and ongoing emissions, the expected increase rate is
0.1-0.3\degree/decade, which would lead to +1.5\degree\ sometime during 2030-2050
if no mitigation measures are taken. Subsequent releases of IPCC
reports have confirmed this conclusion, and strengthened the attribution
of climate change to human influence \cite{ipcc:2021}. The phenomenon
seems to be accelerating, and each of the last four decades has been
successively warmer than any decade that preceded it since 1850.\\

\noindent The IPCC reports that the impacts of anthropogenic climate
change are already perceptible in the intensity and frequency of
climate and weather extremes, such as floods, heat waves and
droughts. These changes affect entire human societies, and concerns
are progressively shifting to adaptation to climate change, especially
in developing countries \cite{ipcc:2022}. Natural ecosystems are also
strongly affected, with an unprecedented collapse in
biodiversity. While land use and sea exploitation are the dominant
drivers of biodiversity loss, climate change is having a growing
impact on biodiversity, exacerbating other drivers. The global
assessment report on biodiversity and ecosystem services of the
Intergovernmental Science-Policy Platform on Biodiversity and
Ecosystem Services (IPBES) states that the rate of species extinction
across the globe is tens to hundreds of times higher than the average
rate over the past 10 million years, and is accelerating
\cite{ipbes:2019}.\\

\noindent The main messages from the IPCC report can be summarized as follows:
(i) every fraction of a degree counts, and there are significant
differences between a +1.5\degree\ and +2.0\degree\ global warming
scenario in terms of adverse consequences; (ii) our choices over the
next two decades are crucial to secure emission pathways remaining
below or only slightly overshooting +1.5\degree; (iii) such pathways require
far-reaching and systemic transitions and transformation in nearly
all sectors of human activity. The latter is emphasized in the
IPBES report, which states that sustainability in our use of nature
for 2030 and beyond may only be achieved through ``fundamental,
system-wide reorganization across technological, economic and social
factors, including paradigms, goals and values''.\\

\noindent The +1.5\degree\ scenario with no or limited overshoot requires
anthropogenic \co\ emissions to drop by 40-60\% by 2030 relative to the
2010 levels \cite{ipcc:2018}. This implies an average reduction rate
of 7.6\% every year over the next decade \cite{unep:2019}. We recently
went through an unexpected episode that illustrates the magnitude of
the necessary changes. In 2020, as a result of the drastic global
response to the COVID-19 pandemic, \co\ emissions fell by 6.4\%
\cite[and subsequently bounced back; see][]{Tollefson:2021}. This is
about the level of reduction we need to achieve every year, sustained
over more than a decade.\\

\noindent Over the past decade, a growing fraction of the scientific
community has recognized the urgency of these results, mirroring the
increasing concern within civil society
\cite[e.g.][]{Achten:2013,LeQuere:2015,Ciers:2019}.
Complementing the legitimate and necessary calls to stand
up as scientists to warn citizens and policymakers of the major threat
posed by climate change \cite{Ripple:2019}, a growing number of
individuals and institutions are now willing to assess the
environmental impact of scientific research as it is currently practiced.\\

\noindent In the field of astronomy and astrophysics, several quantitative
assessments and discussions about our impact on the environment --
mostly focussing on GHG emissions -- have recently been
published. These studies have addressed several different facets of
the community and its activities: a single research institute
\cite{Jahnke:2020}, a national community
\cite{Stevens:2020,VanDerTak:2021}, existing or planned research
infrastructures \cite{Flagey:2020,Aujoux:2021}, high-performance
computing \cite{Portegies-Zwart:2020}, large astronomy meetings
\cite{Burtscher:2020}, and possible mitigating solutions
\cite{Matzner:2019,Burtscher:2021}. Despite their differences, these
studies all find that significant efforts must be made to align our
research practices with IPCC-recommended trajectories for the next
decades.\\

\noindent In this paper, our goal is to contribute to this collective
reflection by presenting a carbon footprint assessment for the
Institut de Recherche en Astrophysique et Plan\'etologie (IRAP)
located in Toulouse, France. Our reporting uses 2019 as the reference
year and includes a large number of emission sources not considered by
earlier studies, e.g. the emissions from purchased goods and services,
and an estimate for the impact of using large-scale research
equipments such as ground-based observatories and space-borne
telescopes and probes. The result suggests that previous efforts were
most likely underestimating the carbon footprint of research in
astronomy by a large factor, which puts us even further away from the
assigned goals for sustainability of human activities. Our work
strengthens the evidence that massive GHG emissions are at the very
heart of our research activities and highlights the urgent need for an ambitious community-wide plan
for the future, challenging the deep cultural roots of astrophysics as
we exercise it today.\\

% Methodology and scope
\section{Methodology and scope}
\label{method}

% Local and national context
\subsection{Local and national context}
\label{method:local}

\noindent In France, the practice of GHG emission assessment in the public and
private sectors was formalized by legislation passed in 2010
\cite{loi:2010}, extended and complemented by further legislation in
2015 \cite{loi:2015}. The objectives and requirements of these GHG
emission assessments are summarized in a practical guide by Ref.
\cite{meem:2016}. Together, these documents identify the private and
public actors that are legally bound to conduct an assessment of their GHG
emissions, and define the minimal standards for conducting, publishing,
and updating such assessments.\\

\noindent A research laboratory, as it is legally structured in France, is not
subject to this legal obligation, which applies to higher-level institutions
such as universities and national research organizations like the ``Centre National de la Recherche Scientifique (CNRS)''. 
The latter, however, are often very large entities, gathering
up to a few tens of thousands employees involved in very different
fields of activity, and spread over a large number of geographical
sites. Conversely, laboratories seem the relevant scale for initiating
such an assessment: they employ a few tens to a few hundreds of people
located at a small number of physical sites, are relatively
homogeneous in terms of their activity, and offer a more direct and
efficient access to the administration, which is key to collecting the
relevant data. \\

\noindent The idea of conducting a carbon footprint assessment at IRAP
was promoted by a group of a dozen persons organized in an official
commission of the institute since 2018. The commission has
representatives on the institute council and its role is to assist the
direction of the institute in all efforts related to reducing the
environmental impact of institute activities, e.g. raising staff
awareness, waste management, promoting environment-friendly options
for travel, etc. The formal decision to conduct a carbon footprint
assessment was taken jointly with the institute's direction, and the
work started in late 2019. Similar developments occurred in six local
institutes that together with IRAP form the ``Observatoire
Midi-Pyr\'en\'ees'' (OMP) research federation. Although these
institutes specialise in different scientific domains (oceanography,
geophysics, biosphere, ecology, climatalogy), it was relevant to join
forces because office space, services and facilities are largely
shared among OMP institutes.

% Methodology and tools
\subsection{Methodology and tools}
\label{method:tools}

\noindent As a first step, 24 OMP staff members, including eight from
IRAP, followed a 40~h training course about how to conduct a carbon
footprint assessment with the \gls{bc} methodology. This training was
funded by the participants' institutes and the CNRS. The training was
delivered by the ``Institut Formation Carbone''\footnote{\url{https://www.if-carbone.com/IFC_WEB}}. \gls{bc} is a carbon
accounting methodology and set of tools that have been developed and
used in France for more than 20 years. It has been applied to private
companies, industries, and structures of all kinds and sizes including
universities and research institutes. The method is compatible with
other reporting methods such as the GHG Protocol \cite{ghg:2004} or
ISO 14064-1 \cite{iso:2018}. The method is based on emission factors
taken from the ``Base Carbone''
database\footnote{\url{http://www.basecarbone.fr/fr/accueil/}} of the French
``Agence pour le D\'eveloppement et la Ma\^itrise de l'Energie
(ADEME)''.\\

\noindent  Our work complements community efforts emerging in France to
systematically track the carbon footprint of research institutes
\cite{ges1point5} by providing an in-depth and more complete
assessment of our institute. This is important because, as we show
later, our results suggest that the perimeter of the assessment needs
to be widened as much as possible in order not to miss potentially
dominant emission sources.\\

\noindent  As recommended in the \gls{bc} methodology, intermediate results were
reported to the institute in July 2021, after approximately six months
of data collection. The purpose of this mid-term restitution is to
inform colleagues, validate the data obtained so far, and agree on
final steps. The present document constitutes the final report, which
will serve as a basis for the definition of a long-term reduction
plan.

% Cartography of activities and fluxes
\subsection{Cartography of activities and flows}
\label{method:carto}

\noindent Carbon accounting following the \gls{bc} method starts with
a cartography of the institute that aims to capture all its activities
and the input and output flows that they generate. The main guiding
principles of this exercise are relevance and completeness. Such a
cartography is meant to provide a strategic view of the institute,
allowing a census of everything that the institute critically depends
upon, as well as a perspective on everything that the institute
produces and delivers to external partners\footnote{This exercise
  goes beyond carbon emissions. It provides a critical examination of
  how an organization's activities depend on fossil fuels and other
  primary resources, thereby revealing the vulnerability of an
  organization to future evolutions in the supply and pricing of
  commodities.}.\\

\noindent The overall philosophy of the \gls{bc} method is to identify the
organization's most powerful lever arms to achieve large GHG emission
reductions globally, rather than compiling a list of the emissions
that an organization is directly or solely responsible for. An
exhaustive approach is key to developing a perennial and effective
reduction plan since it takes into consideration the deep changes that
may be required to achieve significant permanent reductions. Carbon
accounting over a highly restricted scope risks excluding an
organization's dominant sources of emissions and hiding the reasons behind
certain sources of carbon emissions. As such, the reduction measures
based on a restricted carbon footprint assessment may have limited
effectiveness in the long term. As an example, there has been a strong
focus within the scientific community on the impact of air travel, and
several reports have identified it as the major contributor to their
carbon footprint \cite[e.g.][]{Jahnke:2020}. In the case of IRAP --
but the result is likely to apply to many other astronomy institutes
-- we show below that air travel is not the dominant source of
emissions. As such, flights should not be the only target of reduction
measures if IRAP is to achieve a reduction trajectory compatible with
IPCC recommendations.\\

\noindent  In this cartography, the IRAP is described as a structure where the following core activities and duties are performed:
\begin{enumerate}
\item Instrument development including hardware and software
\item Observations, laboratory experiments, and data analysis
\item Analytical and numerical modelling of natural phenomena
\item Teaching, training, and public outreach
\item Animation and participation in the scientific community
\end{enumerate}
The fifth category includes a broad range of activities, such as
participation in national and international conferences, expert
panels, selection committees and time allocation committees at the
national and international level. The fourth category extends
significantly beyond the perimeter of the institute.  We decided to
mostly exclude it from our reporting since most of the impact of
teaching and training, including the commuting of students, is more
appropriately assessed at the level of universities and schools. What
remains inside our scope from the fourth category are expenses
connected to students directly using our resources and facilities
(e.g. electricity used during their internships in the institute), the
commuting of students affiliated to IRAP for the whole of 2019 (mostly
PhDs), and regular transfers of the staff between the institute
facilities and teaching sites.\\

\noindent  The above activities are associated with flows that can be inputs,
outputs, or internal in nature. These can be further organized into
the following categories: (i) people, e.g. commuting of the staff,
scientific visitors, professional travel, conference attendees, etc ;
(ii) material, e.g. electronic components, chemical products, computer
equipments, etc ; (iii) services, e.g. subcontracting, numerical
resources, observational data, etc ; (iv) internal support,
e.g. heating, electricity, administration, canteen, etc.\\

\noindent A major input to our activity is the large amounts of observational
data from research infrastructures like ground-based observatories or
space-borne telescopes. This is counted as external service, that we
generally do not own, control or pay for, but nevertheless has an
significantly impact on our activities. Similarly, some crucial
numerical resources such as supercomputing, cloud storage,
astronomical databases and various internet facilities, are counted as
external services because their operations and related impacts are
external to the institute (computing and data storage performed on
site at IRAP is accounted for in the purchase of computer equipment
and consumption of electricity however). We note that emissions in the
same flow category are not necessarily treated in the same way.  In
practice, the accounting is driven by the format and availability of
the related information.

% Scope and boundaries
\subsection{Scope and boundaries}
\label{method:scope}

\noindent Based on the above cartography of our activities and
associated flows, we define the temporal, organizational, and
operational scope of our carbon footprint assessment as follows:\\

\textbf{Reference year}: We adopted 2019 as reference year. As this
was before the COVID-19 pandemic, it presents a snapshot of our recent
activities as they were before lockdowns and travel
restrictions. Whether the carbon footprint will be affected in the
long term by changes in activity or attitude patterns following
COVID-19 remains to be seen. Moreover, the actual distribution of our
emissions, as presented below, suggests that that COVID-related
restrictions in 2020 and 2021 had only a minor impact on our carbon
footprint.\\

\textbf{Organisation}: The assessment concerns IRAP, a research
institute spread over three sites (two in Toulouse and one in Tarbes),
two of which are shared with other institutes (the Belin site in
Toulouse and the Tarbes site). These three buildings and their related
support services are areas over which IRAP has significant effective
control, and where it can rapidly implement mitigation measures. Our
footprint goes beyond these sites, and the numerous facilities
that are necessary to our daily work, such as external computing
centers and other research infrastructures, are included as external
service providers in our assessment. Teaching and training activities,
 which concerns a non-negligible fraction of IRAP staff, is essentially
left out of the scope as explained above.\\

\textbf{People}: We consider all persons that were employed for the
whole of 2019, whether their status was permanent or not. This
excludes most undergraduate students, short-term interns, and PhDs and
postdocs that left the institute during 2019. In total, we identified
116 researchers, 28 postdocs, 78 engineers, technicians and
administrative staff, and 41 PhD students that were employed at IRAP
over the full year of 2019.\\

\textbf{Operations}: According to the ISO14069 nomenclature, we
include emissions for the following categories of activity (usually
termed scope): direct emissions from owned or controlled sources
(scope 1), indirect emissions from the generation of purchased energy
(scope 2), and all other indirect emissions (scope 3). In our case,
the latter category is by far the dominant one.

\begin{table*}[pt!]
\centering
\begin{tabular}{lcc}
\hline
Source & Activity data & Emission data (\tco)  \\
\hline
Electricity & Roche: 1863 MWh & $113 \pm 11$ \\
& Belin: 398 $\pm$ 80 MWh & $24 \pm 5$ \\
& Tarbes: 15  $\pm$ 3 MWh & $1 \pm 0.2$ \\
& Total: 2276  $\pm$ 80 MWh & $ 138 \pm 12$ \\
\hline
Heating & Roche: 857 MWh &  $59 \pm 18$ \\
& Belin: 177  $\pm$ 35 MWh & $40 \pm 8$ \\
& Tarbes: 38 $\pm$ 8 MWh & $9 \pm 2$ \\
& Total: 1072 $\pm$ 36 MWh & $108 \pm 20$ \\
\hline
Water & Roche: 4271 m$^3$ & $1.7 \pm 0.3$ \\
& Belin: 428 $\pm$ 86 m$^3$ & $0.2 \pm 0.04$ \\
& Tarbes: 45 $\pm$ 9 m$^3$ & $0.01 \pm 0.002$ \\
& Total: 4744 $\pm$ 86 m$^3$ & $2 \pm 0.3$ \\
\hline
Air conditioning & Roche: 11.1 kg of R410A & $21 \pm 6$ \\
 & Tarbes: 1.59 kg of R410A & $3.1 \pm 0.9$\\
 & \hspace{24px} 0.92 kg of R22  & $1.6 \pm 0.5$ \\
 & \hspace{24px} 0.19 kg of R32 &$ 0.13 \pm 0.04$ \\
  & Total & $26 \pm 6$ \\ 
\hline
Waste & $155 \pm 41 $ tons& $55 \pm 20$ \\
\hline
Food & $44500 \pm 15000$ meals  & $85 \pm 50 $ \\
\hline
Commuting & $(1.6 \pm 0.3) \times 10^6$\km&  $174 \pm 67$ \\
\hline
Internal commuting & $(1.0 \pm 0.2)  \times 10^5$\km &  $10 \pm 4$ \\
\hline
Professional travel & flight : $(5.9 \pm 0.1)  \times 10^6$\km & $1126 \pm 48 $ \\
 & train: $(2.5 \pm 0.03)  \times 10^5$\km & $1.2 \pm 0.1$ \\
 & car/cab: $(1.8 \pm 0.06) \times 10^5$\km & $42 \pm 6.0$\\
 \hline
Hotel & $3996 \pm 59$ nights & $75 \pm 6$ \\
\hline
Computer equipment  & 139 (139-153) units & $81 \pm 40$ \\
\hline
Goods and services  & 3.657\,M\euro & $1335 \pm 342$ \\
\hline
External computing & $7.0 \pm 3.5$ MhCPU & $33 \pm 26$ \\
\hline
External storage &  $293 \pm 129$ TB & 26 (4-63) \\
\hline
Data flow & $293 \pm 129$ TB  & 1.5 (0.3-3.2) \\
\hline
Observational data & space: 46 missions & $2800 \pm 600 $ \\
 & ground: 39 observatories & $1300 \pm 500 $ \\
\hline
\multicolumn{2}{l}{Total (restricted perimeter)} & \gestot $ \pm 359$ \\
\multicolumn{2}{l}{Total (full perimeter)} & \gestotxl $ \pm 860$ \\
\hline
\end{tabular}
\caption{Summary of emission sources. Uncertainties are quoted as a range of values when they are strongly
  asymmetric. The uncertainties for the emission data include relative
  uncertainties on activity data and emission factors added in
  quadrature. For some categories (meals, commuting, internal
  commuting, external storage), activity data are presented in an
  aggregate form and uncertainties are propagated from individual
  items quadratically. The total is given over the full perimeter of
  the assessment, and over a restricted perimeter that excludes the
  contribution of external research infrastructures including external
  computing and storage.}
\label{tab:summary} 
\end{table*}

% Data collection and emissions assessment
\section{Data collection and emissions assessment}
\label{res}

% User survey
\subsection{User survey}
\label{res:survey}

\noindent Most of the data used in the assessment were obtained from our
administration in the form of listings of our consumption, purchases,
and travel. For the data that were not available to the administration
(commuting practices, meals, and use of computing/data storage
resources external to our institute), we conducted an online user
survey. The survey form was based on a protocol provided in
\cite{ges1point5}, but was expanded and adapted to suit our needs.\\

\noindent  The response rate was 59\% among faculty and staff researchers, 51\%
among engineering, technical, and administrative personnel, and 29\%
among PhD students and postdocs. The lower response rate among PhD
students and postdocs is due to the fact that many of those that were
at our institute in 2019 had left by the time we conducted the
survey. We extrapolated the data collected to the entire population by
assuming that the sample available is representative within each of
the three personnel subcategories listed above. Following the
\gls{bc} methodological guide we assigned to these data a nominal
uncertainty of 50\%.\\

% Electricity and heating
\subsection{Electricity and heating}
\label{res:heatelec}

\begin{table*}[t!]
\centering
\begin{tabular}{ l  c  c  c  c  c  c  c }
\hline
\hline
Source & Incoming  & Delivered& \multicolumn{2}{c}{Combustion only} &  \multicolumn{2}{c}{Full LCA} &   \\
& MWh PCI & MWh & $\frac{\rm t CO2e}{\rm MWh \, PCI}$ & \tco & $\frac{\rm t CO2e}{\rm MWh \, PCI}$ & \tco \\
\hline
Wood &45927&&0.0132&606&0.0244&1121\\
Natural gas &10585 && 0.187 & 1979 &0.227&2403\\
\hline
Total & 56512 & 51136 & & 2585 & & 3524\\
EF ($\frac{\rm kg CO2e}{\rm kWh}$) & & & & 0.0506& & \textbf{0.0689}\\
\end{tabular}
\caption{Calculation of the emission factor (EF) of the university boiler house in 2019, based on the total wood and gas consumption during the year. The final emission factor is obtained by dividing the total CO$_{\rm 2e}$ emissions obtained from life cycle analysis (LCA) by the total energy delivered by the boiler house in 2019.}
\label{tab:EFsge} 
\end{table*}

\begin{figure}[!t]
\begin{center}
\includegraphics[width=0.7\textwidth]{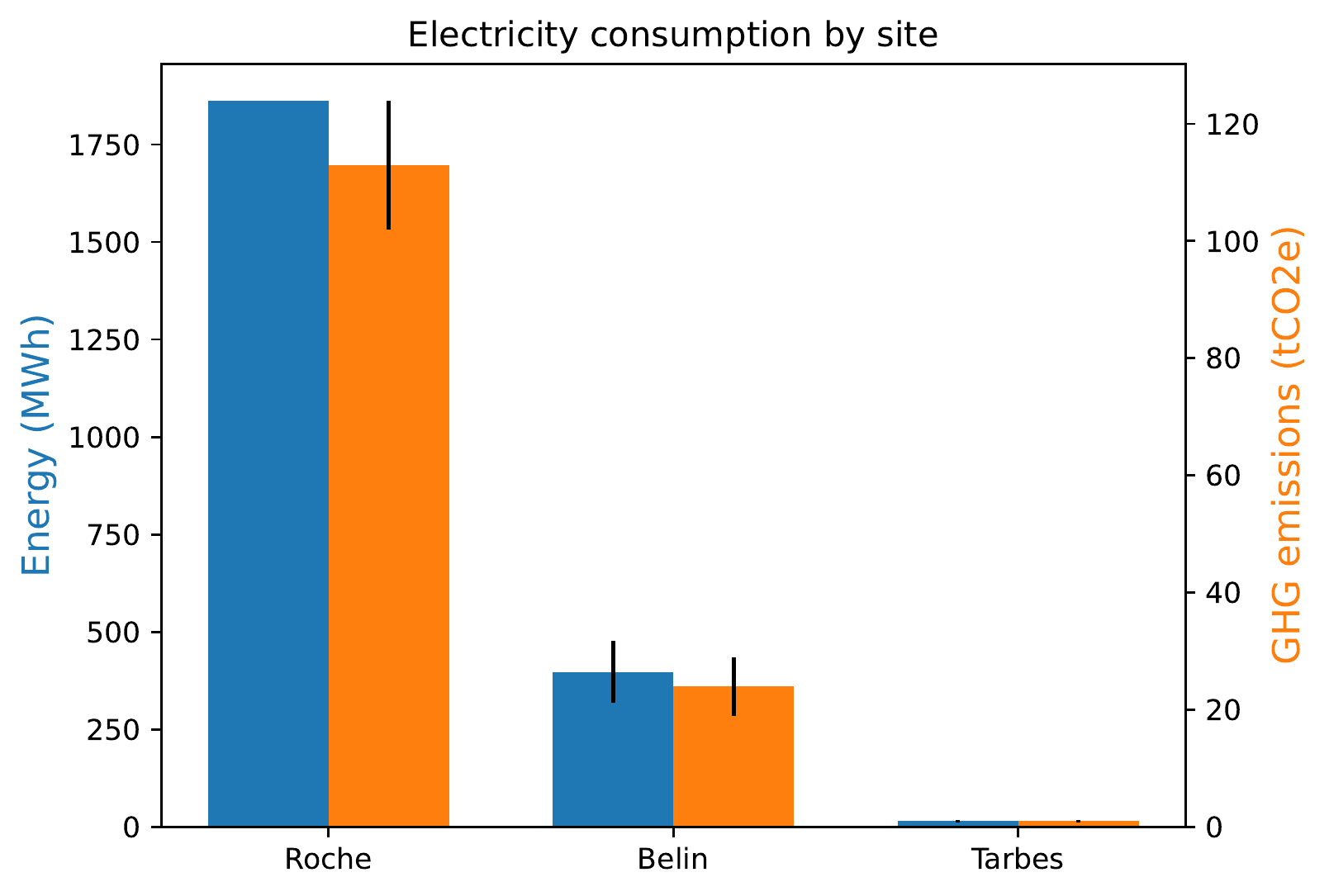} 
\includegraphics[width=0.7\textwidth]{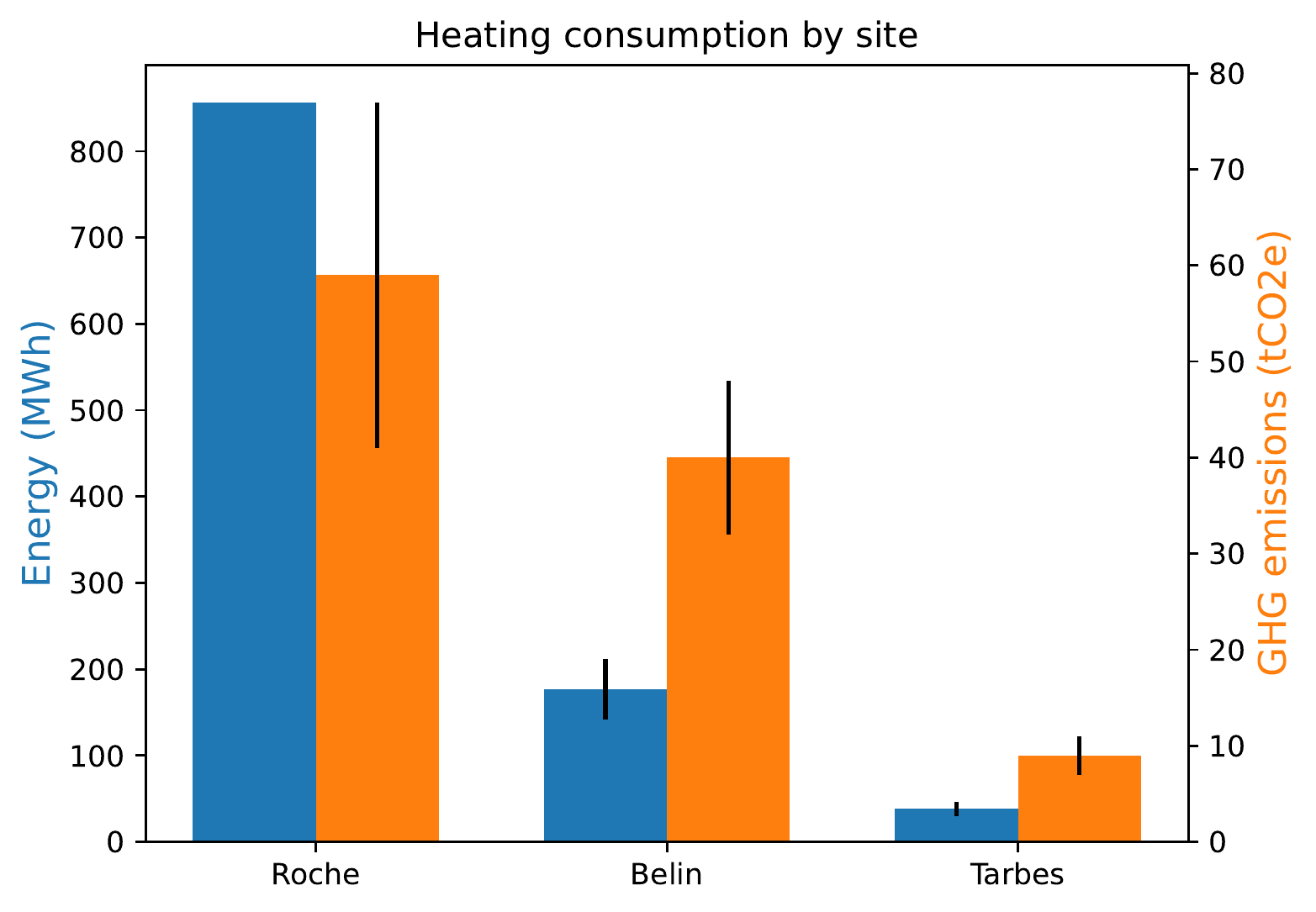} 
\caption{Electricity and heating consumption at IRAP, and the
  corresponding GHG emissions. Data for the three sites of the
  institute are shown separately.}
\label{fig:elecheat}
\end{center}
\end{figure}

\noindent As described in Sect. \ref{method:scope}, IRAP is spread over three
different sites: the Roche and Belin buildings in Toulouse, and
another building in the city of Tarbes. The Roche building is entirely
occupied by IRAP while the Belin and Tarbes premises are shared with
other laboratories. The Roche site is connected to the heating network
of the university, while the heating on the Belin and Tarbes sites is
provided by natural gas boilers. Since the electricity and gas consumption
is measured at the building level, we estimated our share at the
Belin and Tarbes sites based on the relative surface occupancy of the
different laboratories. Emission factors for the electricity and gas
were taken from ADEME.\\

\noindent The university heating network is supplied by a boiler house fuelled
by both wood and natural gas. The main fuel in 2019 was wood. The
preliminary emission factor for the 2019 heat production provided by
the boiler house services did not include the upstream and methane
emissions for the use of wood as a fuel. We therefore calculated our
own emission factor based on a life cycle analysis (LCA) for wood and gas
fuels and the heat production data of the boiler house in 2019, and
obtained a final emission factor of 0.0689\kgcoperkwh. Table
\ref{tab:EFsge} provides details on the calculation of this emission
factor. Note that wood combustion is considered to emit biogenic
CO$_2$ not fossil CO$_2$. Whatever its origin, biogenic or fossil,
CO$_2$ gas increases the greenhouse effect in the same
way. However, unlike fossil fuels, biomass can be renewed on a human
timescale, although the exact length of the cycle varies considerably
(e.g. annual crops vs. forests). For biomass of forest origin, biogenic
carbon emissions are not counted if this biomass comes from a country
where harvesting remains below the biological growth of the forest,
which is the case in France. Therefore, the emission factor for wood
combustion listed in Table \ref{tab:EFsge} only takes into account
methane emissions.\\

\noindent For the natural gas used in the heating of the Belin and Tarbes sites,
we use an emission factor of 0.227\kgcoperkwh corresponding to the
French gas mix in 2015 (more recent estimates were not available). For
the electricity, we use an emission factor of 0.0607\kgcoperkwh
corresponding to the French electricity mix in 2019.\\

\noindent Electricity and heating consumption, together with their corresponding
GHG emissions, are listed in Table \ref{tab:elec_heat} and presented
in Fig. \ref{fig:elecheat}. We note that there are no uncertainties
associated with the electricity and heat consumption data for the Roche site
as it is accurately metered. For the Belin and Tarbes sites, we added a conservative
$20\%$ uncertainty related to the calculation of IRAP's share based on surface occupancy.
In addition, we used uncertainties of $5\%$
and $10\%$ for the natural gas and electricity emission factors,
respectively, as provided in the ADEME database. For the university
heating network, we used the generic $30\%$ applied to heating networks
in the ADEME database. In total, electricity and heating consumptions
emit $246 \pm 23$\tco.\\

% Waste production
\subsection{Waste production}
\label{res:waste}

\noindent Waste production is associated with a carbon footprint that was
estimated from the information provided by the logistical services of
the institute. The weight of IRAP's waste is not systematically
calculated, so our estimates are based on the number of containers for
each type of waste, the frequency of collection, and the volume of
containers. Assuming 41 weeks with full containers (a time period
corresponding to the university's operations excluding vacations), we
estimated a total volume of waste of $508$\,m$^3$ per year, separated
into several types: paper ($9\%$ of total volume), cardboard ($17\%$),
dry recyclable waste ($16\%$), household waste ($58\%$). Assuming a
density of 0.3\,kg/l for dry waste, we deduce a yearly collected
weight of $155$ tons. In the absence of a better estimate, we assumed
an uncertainty of 50\% on the collected mass of each type of waste for
each of the three sites of the institute.\\

\noindent  Appropriate emission factors for each type of waste, which depends on
their end of life, i.e. burning or recycling, were taken from the
ADEME database with a recommended uncertainty of 30\%. This leads to
a total emission from waste production of $55$\tco, with a final
uncertainty of 20\tco resulting from quadratic combination of relative
uncertainties for each type of waste and quadratic sum over all types
of waste.\\

\noindent IRAP's waste-related carbon footprint is essentially due to household
waste, $58\%$ of total emissions, that is burnt, and to the recyclable
waste, $39\%$ of emissions, that is only partially
recycled. Increasing the share of waste that is recycled, and most
importantly limiting the amount of waste produced, are necessary to
reduce the carbon footprint of IRAP's waste production.

% Water consumption
\subsection{Water consumption}
\label{res:water}

\begin{table*}[t!]
\centering
\begin{tabular}{ l c c c c c c c c }
\hline
\hline
Site & \multicolumn{2}{c}{Electricity} & \multicolumn{2}{c}{Heating} & \multicolumn{2}{c}{Water}  \\
& MWh &\tco & MWh &\tco & m$^{3}$& \tco\\
\hline
Roche &1863 & 113 $\pm$ 11& 857 &59 $\pm$ 18 & 4271 & 1.7 $\pm$ 0.3\\
Belin & 398 $\pm$ 80 & 24 $\pm$ 5 &177 $\pm$ 35 &40 $\pm$ 8 & $428 \pm 86$ & 0.2 $\pm$ 0.04 \\
Tarbes &15 $\pm$ 3 &1 $\pm$ 0.2& 38 $\pm$ 8 &9 $\pm$ 2 & $45 \pm 9$ & 0.01 $\pm$ 0.002\\
\hline
Total & 2276 $\pm$ 80 & 138 $\pm$ 12&1072 $\pm$ 36 &108 $\pm$ 20 & $4744 \pm 86$ & 2 $\pm$ 0.3\\
\end{tabular}
\caption{Electricity, heating and water consumption per site and their corresponding GHG emissions.}
\label{tab:elec_heat} 
\end{table*}

\noindent  Water consumption is measured at the building level. As for
electricity and heating, we estimated our share of the water
consumption at the Belin and Tarbes sites based on the relative
surface occupancy of the laboratories. We associated a conservative
$20\%$ uncertainty on these estimates, except for the Roche site where
water consumption is metered and should have a negligible uncertainty.\\

\noindent  The related GHG emissions were estimated from the ADEME emission
factors for water distribution $0.262 \pm 0.052$ \kgcoperm and
wastewater treatment $0.132 \pm 0.015$ \kgcoperm. Water consumption is
detailed in Table \ref{tab:elec_heat} and contributes to a total of $2
\pm 0.3$ \tco.

% Air conditioning
\subsection{Air conditioning}
\label{res:aircond}

\noindent In addition to the electricity consumption already taken
into account in Sect. \ref{res:heatelec}, air conditioning also
contributes to GHG emissions due to refrigerant gas leakage. \\

\noindent For the Roche site, the leakage was estimated based on the
amount of gas refilled during the year, totalling 11.1 kg of R410A
gas. Using a global warming potential (GWP) of $1928$ over 100\,yr, with a $30\%$
uncertainty as recommended by ADEME, this corresponds to
$21\pm6$\tco. The uncertainty is here related to the GWP only since
the amount of gas refilled is precisely monitored and billed by the
maintenance company.\\

\noindent  For the Tarbes site, the IRAP share of refrigerant gas leakages are
0.92 kg of R22, 0.19 kg of R32, and 1.59 kg of R410A. Using global
warming potentials of 1760 and 677 for R22 and R32 respectively, we
obtain equivalent emissions of $4.8\pm1.0$\tco assuming $30\%$
uncertainty on the GWP.\\

\noindent Unfortunately, we could not obtain the refrigerant gas consumptions
for the air conditioning system on the Belin site. However, air
conditioning is mainly used in technical and common rooms in this
building, unlike the Roche site where air conditioning is provided in
each office, suggesting that GHG emissions due to refrigerant gas
leakage on the Belin is significantly below the result for the Roche
site. The final IRAP carbon footprint should not be significantly
impacted by this omission. Overall, refrigerant gas leakage
contributes a total of $26\pm6$\tco.

% Meals
\subsection{Meals}
\label{res:meals}

\noindent Meals taken at the workplace are necessary to the personnel to carry
out their work, and therefore must be included in a GHG
assessment. Following the ADEME recommendations for entities that are
not part of the food and agriculture business, we take a simplified
approach based on the number of meals \cite{ademe-repas}. The main
factor that determines GHG emissions from a meal is its content in
animal products\footnote{See, e.g.,
  \url{https://ourworldindata.org/environmental-impacts-of-food}.}.\\

\noindent From our user survey, we estimated the number of meals in three
categories: standard, flexitarian (reduced amount of animal products),
and vegetarian. As for all activity data extracted from the user
survey, we assume these numbers have an uncertainty of 50\%. We apply
the emission factors proposed by ADEME to these three meal categories
under the hypothesis that the meat content in non-vegetarian meals is
composed of 25\% of high-carbon meats (namely beef) and for 75\% of
lower-carbon meats (namely chicken) or fish. This yields
2.585 \kco/meal for standard meals, 1.103 \kco/meal for flexitarian
meals, and 0.510 \kco/meal for vegetarian meals, with uncertainties of
50\%.\\

\begin{figure}[t!]
\begin{center}
\includegraphics[width=0.7\textwidth]{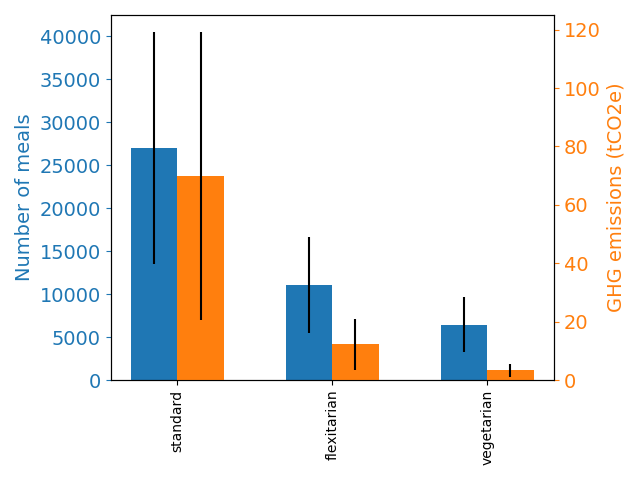} 
\caption{\label{fig:meals} Number of meals and their corresponding GHG emissions.}
\end{center}
\end{figure}

\noindent  The results for the number of meals and GHG emissions are shown in
Fig.~\ref{fig:meals}. Meals contribute a total of $85 \pm 50$~\tco to
IRAP's carbon footprint. We note that 75\% of these meals are provided
by a canteen, and, specifically, 73\% by the CNRS canteen.\\

% Commuting
\subsection{Commuting}
\label{res:commute}

\begin{figure}[!t]
\begin{center}
\includegraphics[width=0.7\textwidth]{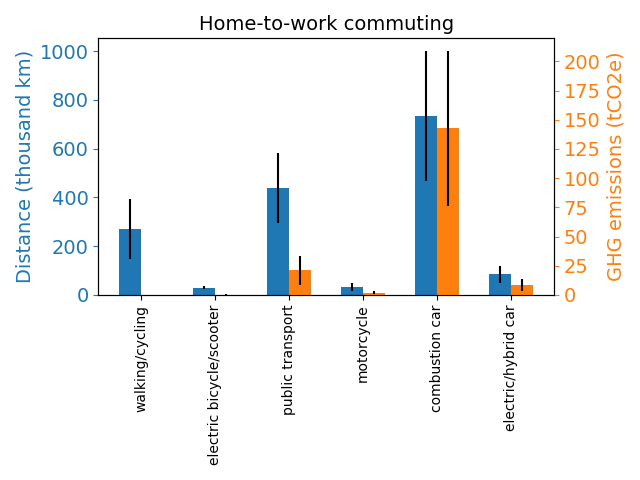}
\includegraphics[width=0.7\textwidth]{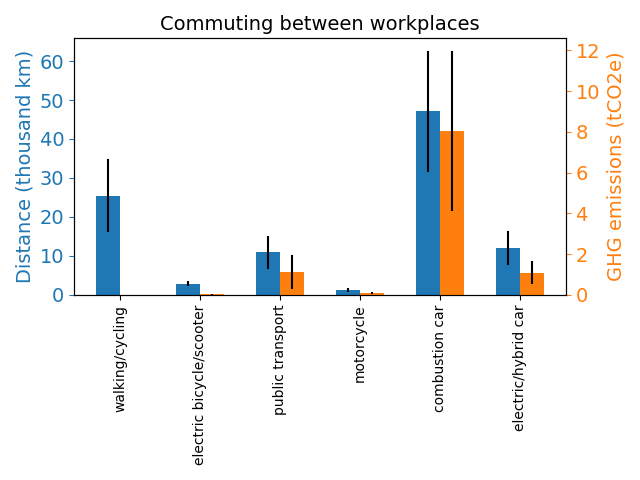}
\caption{\label{fig:commuting} Distances and corresponding GHG
  emissions from home-to-work commuting and commuting between
  workplaces. Distances and GHG emissions are shown according to the
  mode of transportation. We used several subcategories and accounted
  for ride-sharing to estimate the GHG emissions, see text for
  details. When aggregating distances for several subcategories,
  uncertainties are propagated quadratically.}
\end{center}
\end{figure}

\noindent  Emissions from home-to-work commuting are included in IRAP's GHG
assessment according to French legislation and the \gls{bc} method. We
also consider here commuting between workplaces (e.g., between the
institute and a teaching site). These trips are not covered in our
assessment of professional travel since they are undertaken by IRAP
personnel via private means. We estimated commuting distances using
the survey (uncertainties are assumed to be 50\%) and converted them
to GHG emissions using \gls{bc} tools. The modes of transportation
that we consider, along with their emission factors, are listed in
Table~\ref{tab:fe_transport}. Walking and cycling are assumed to
generate zero emissions. We also took into account ride-sharing for
cars and motorcycles. The results are summarized in
Fig.~\ref{fig:commuting}.\\

\begin{table}[!p]
\begin{center}
\rotatebox{90}{
\begin{tabular}{lccc}
\hline
Transportation mean & Fabrication & Upstream & Combustion \\
\hline
Petrol car & 0 & 37 \gco/km $\pm 60\%$ & 164 \gco/km $\pm 60\%$  \\
Diesel car & 0 & 39 \gco/km $\pm 60\%$ & 151 \gco/km $\pm 60\%$ \\
Electric car & 84 \gco/km $\pm 70\%$ & 20 \gco/km $\pm 70\%$ & 0 \\
Hybrid car & 48 \gco/km $\pm 70\%$ & 29 \gco/km $\pm 70\%$ & 106 \gco/km $\pm 70\%$ \\
Motorcycle & 0 & 12 \gco/km $\pm 60\%$ & 52 \gco/km $\pm 60\%$ \\
Bus & 0 & 0  & 137 \gco/km/pass. $\pm 60\%$\\
Subway/Tram & 0 & 3.0 \gco/km/pass. $\pm 60\%$ & 0 \\
Train & 0 & 2.5 \gco/km/pass. $\pm 60\%$ & 0\\
Electric bike/scooter & 0 & 11 \gco/km $\pm 50\%$ & 0 \\
\hline
\end{tabular}
}
\caption{\label{tab:fe_transport} Individual modes of transportation
  considered in our estimate of GHG emissions from commuting with
  their emission factors and uncertainties. Emission factors are
  decomposed into fabrication of batteries (electric and hybrid cars
  only), upstream emissions related to the production of fuel/energy,
  and combustion. For public transportation, emission factors are given per passenger.}
\end{center}
\end{table}

\noindent  The total estimated GHG emissions are $174 \pm 67$~\tco from
home-to-work commuting and $10\pm4$~\tco from commuting between
workplaces. In both cases, conventional cars represent the principal
source of emissions. We note that for commuting between workplaces the
median distance travelled by car is 2.5~km.\\

\noindent We also evaluated the impact of working remotely on GHG
emissions. In 2019 (i.e. before the COVID-19 pandemic), working
remotely was a relatively rare practice at our institute. Through the
survey we estimated that remote working was practised by staff members
for a grand total of about 100 days per week, i.e., less than 10\% of
the total working time. This permitted to avoid 31~\tco from
commuting. However, working remotely is also associated with rebound
emissions that we estimated based on Ref. \cite{ademe-teletravail}. The
rebound effects taken into account are those due to new transport
usage, the usage of the private residence as workplace (increased
heating and energy consumption etc.), and increased use of video
communications. This yields $\sim$9~\tco of rebound emissions, and
therefore a net balance of 22~\tco that were avoided in 2019 thanks to
remote working. We emphasise that this figure should not to be
subtracted from our final total. We evaluate them here simply to
provide an order of magnitude estimate for the reductions that would be 
made possible by remote working.

% Professional travels and hotel stays
\subsection{Professional travel and hotel stays}
\label{res:travels}

\begin{figure}[!t]
\begin{center}
\includegraphics[width=0.7\textwidth]{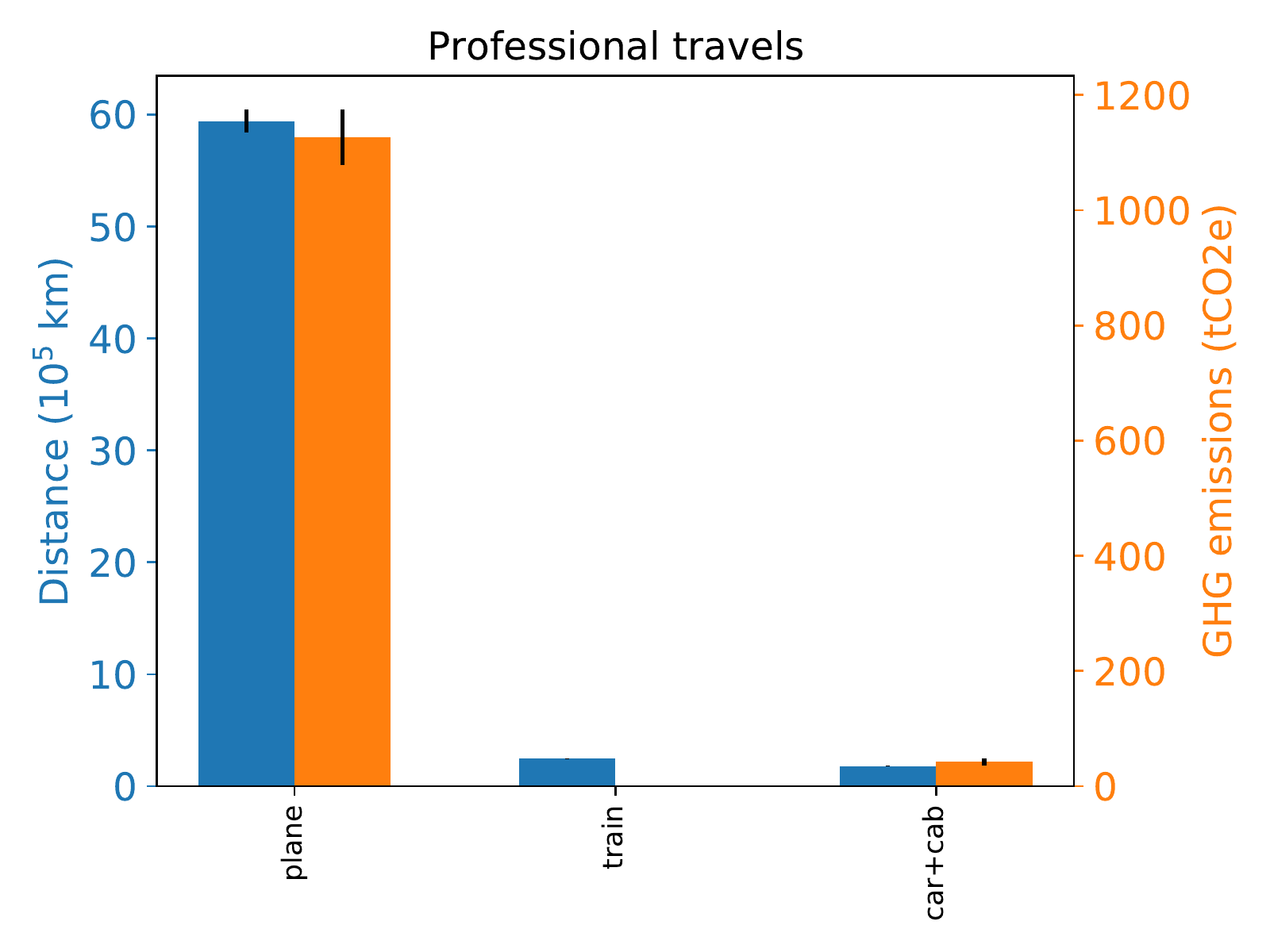} 
\caption{Distances and GHG emissions from professional
  travel, shown according to the mode
  of transportation, with car and taxi grouped into a single
  category.}
\label{fig:missions} 
\end{center}
\end{figure}

\noindent  The emissions related to professional travel, transport and hotel
nights, were estimated based on travel listings containing the
following information: transport mode (flight, train, car...),
departure and destination locations, and dates. For each trip, online
tools were used to estimate the distance from departure to
destination\footnote{\url{https://labos1point5.org/ges-1point5}}. We
assumed an uncertainty of $20\%$ on travelled distances, except for
flights outside France with an increased uncertainty of $50\%$ to
include the non-geodesic distances induced by possible flight
connections.\\

\noindent  GHG emissions were computed for each trip based on transport mode and
distance, and the appropriate emission factor retrieved from the ADEME
database. Regarding air travel emission factors, three different
emission factors are provided depending on travelled distance, and the
effect of contrails is taken into account by multiplying the final \co
equivalent emissions by a factor two, as recommended by
ADEME. Uncertainties on emission factors were taken from the ADEME
database, with in particular a 45\% uncertainty for flights, and we
combined them quadratically with the relative uncertainty on travelled
distances. The total distances travelled by IRAP personnel per transport
mode in 2019 (see also Fig. \ref{fig:missions}), and the resulting \coe
emissions are presented in Table \ref{tab:missions}.\\

\begin{table*}[!t]
\centering
\begin{tabular}{ccccc}
\hline
Transport mode & Distance (km) & Emissions (\tco) \\
\hline
Flight &  5942809 ($93.3\%$) & 1126.2 ($96.3\%$) \\
Train & 247203  ($3.9\%$) & 1.2 ($0.1\%$) \\
Car & 179794 ($2.8\%$) & 41.9 ($3.6\%$) \\
Cab & 728 ($0.01\%$) & 0.2 ($0.01\%$) \\
\hline
\end{tabular}
\caption{Total distances and associated emissions from professional
  trips by IRAP staff members for the different modes of
  transportation. Their relative contribution is shown in
  parentheses. Uncertainties on the figures can be found in Table \ref{tab:summary}.}
\label{tab:missions} 
\end{table*}

\noindent  The professional travel considered here includes trips by
people who are not employed at IRAP (e.g. visitors for short-term
collaboration, seminar speakers,...). These are included because they
are part of the normal working of the institute, and a necessary
contribution to our research activities. They typically consist in one
return trip per person, and represent about 15\% of the carbon
emissions due to professional travel at IRAP.\\

\noindent The total distance travelled in France is $1.05 \times
10^6$\km, about the same distance in Europe, and $4.27 \times 10^6$\km
outside of Europe. The average distance travelled per return trips are
1000, 2100, and 15000\km, respectively. The average distance travelled
per person per year is about 25000\km, or 34000\km\ if administrative
and technical staff are excluded from the calculation (i.e. assuming
that trips are only undertaken by research staff). This corresponds to
$\sim4.6$\tco and $\sim6.3$\tco per person per year,
respectively. Train represents $3.8\%$ of the travelled distance, and
$58\%$ of ground transport, but only $0.1\%$ of emissions due to its
very low emission factor compared to aircrafts and cars.\\

\begin{figure}[!t]
\begin{center}
\includegraphics[width=0.7\textwidth]{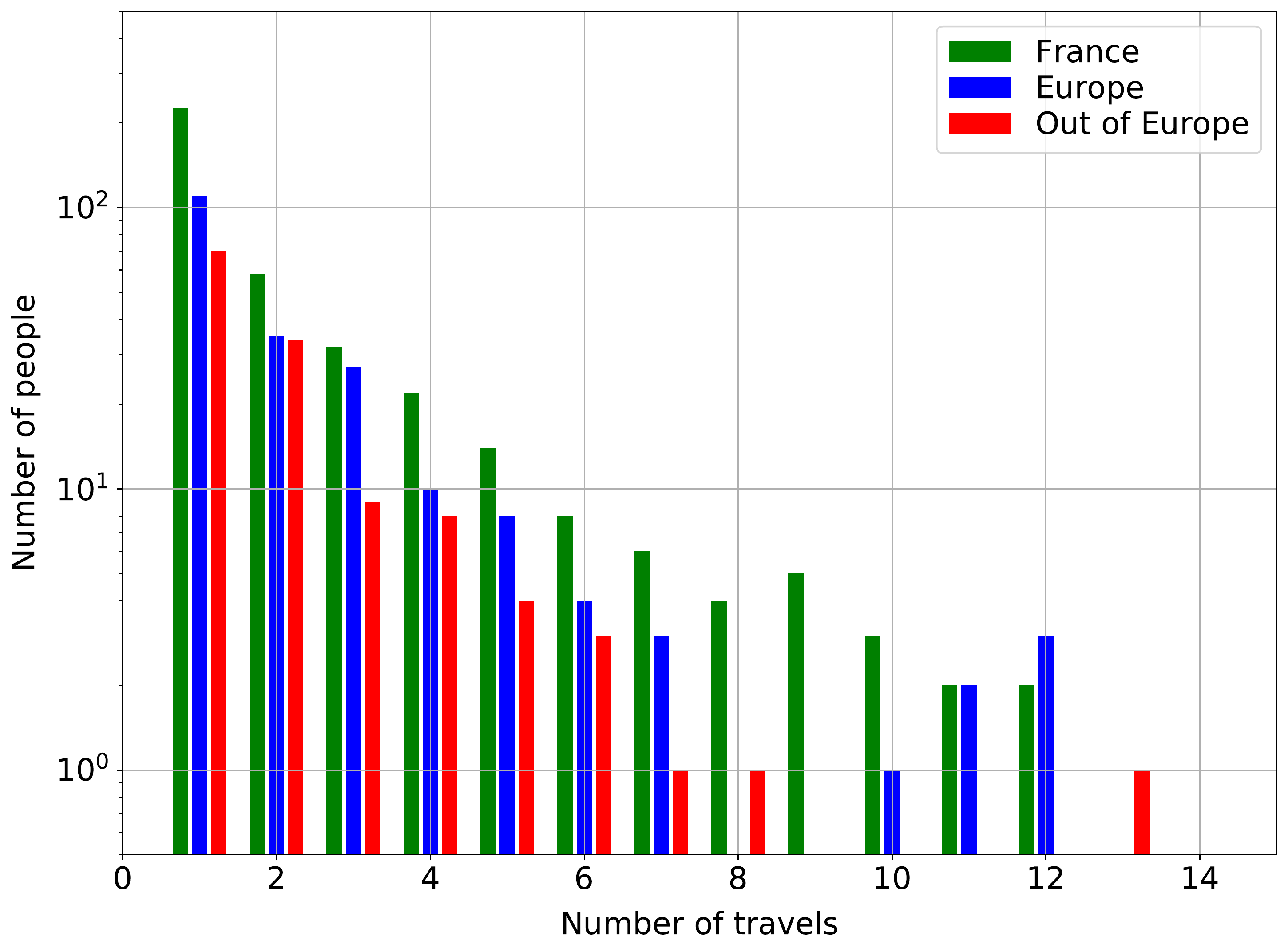}
\includegraphics[width=0.7\textwidth]{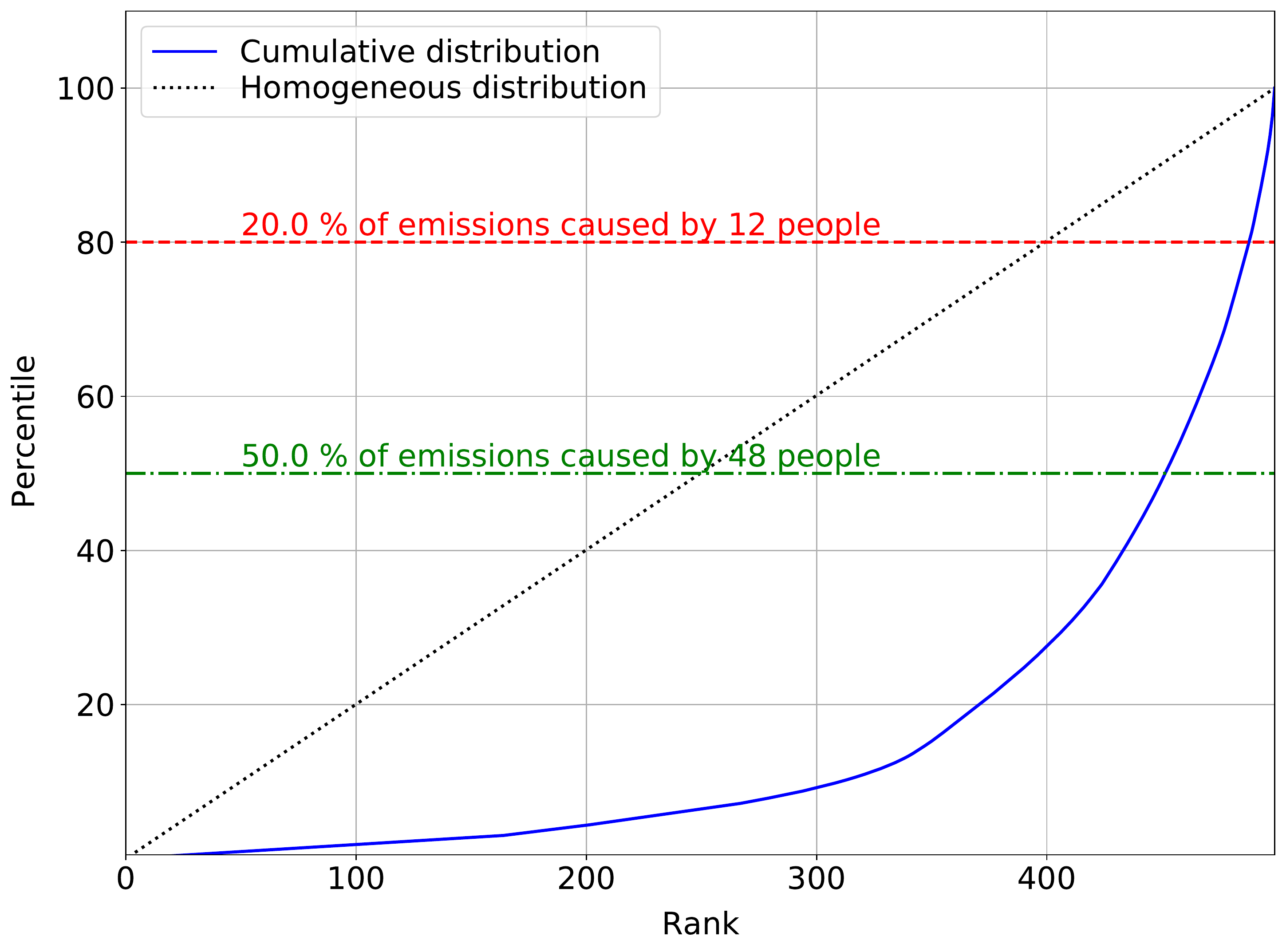}
\caption{Top panel: Distribution of the number of trips per person per
  year as a function of destination (France, Europe, out of
  Europe). Bottom panel: Cumulative distribution of GHG emissions over
  travellers (including IRAP staff and visitors).}
\label{fig:traveldistrib}
\end{center}
\end{figure}

\noindent  The professional travel practices and the associated emissions are
very unevenly distributed within the institute, as illustrated in
Fig.~\ref{fig:traveldistrib}. About 20\% of IRAP's professional travel
emissions are due to only a dozen staff members, and 50\% of the
emissions originates from fewer than 20\% of the travellers. Roughly
$10$\% of IRAP staff members did not travel at all in 2019. In principle, this
concentrated emission profile should make it easier to engage efficient
reduction measures, and reasonable limits on the number of allowed
trips per year would translate into significant emission
reductions. This is addressed more quantitatively in
Sect.~\ref{discu:reduc}.\\

\noindent  The duration and destination of each trip yields the number of nights
away from home, assumed to be spent in hotels, with a $20\%$
uncertainty on that number to cover possible inaccuracies in the
listings or alternative accommodation (e.g. with friends or
family). Emission factors for hotel nights in each country were taken
from the full set of conversion factors proposed by the United Kingdom
Department for Business, Energy and Industrial
Strategy\footnote{https://www.gov.uk/government/publications/greenhouse-gas-reporting-conversion-factors-2019},
with associated uncertainties of $80\%$. The total number of hotel
nights was $3996$ in 2019, leading to $75.2$\tco emissions with an
uncertainty of $6.4$\tco resulting from quadratic combination of
relative uncertainties for each travel and quadratic sum over all
missions.\\

\noindent  In summary, professional travel emissions contributed $1245$\tco\ of
IRAP's GHG emissions in 2019, with 94\% of that amount arising from
transport, mostly air travel, and a non-negligible 6\% arising from
hotel accommodation.

% Purchase of goods and services
\subsection{Purchase of good and services}
\label{res:purchase}

\begin{table*}[t!]
\centering
\begin{tabular}{ccccc}
\hline
ID & Category & Expense & Emission & Share \\
 &  &  (k\euro) & (\tco) & (\%) \\
\hline 
E & Consulting/Insurance/Human Resources &  493 & 122 & 9.17 \\
I & Computing/Telecommunications &  668 & 117 & 8.77 \\
O & Optical &  518 & 174 & 13.01 \\
T & Electronics &  673 & 262 & 19.64 \\
R & Mechanics/Automation &  354 & 144 & 10.78 \\
C & Communication/Documentation &  159 & 44 & 3.30 \\
P & Nuclear/Particle Physics &  144 & 134 & 10.02 \\
A & General supplies &  139 & 59 & 4.43 \\
F & Freight/Transport &  43 & 28 & 2.16 \\
V & Vacuum &  82 & 55 & 4.10 \\
B & Buildings/Infrastructure &  128 & 53 & 3.96 \\
G & Cryogenics/Laboratory gases &  28 & 28 & 2.10 \\
N & Chemistry/Biology &  47 & 54 & 4.05 \\
\hline
\end{tabular}
\caption{Distribution of the main sources of GHG emissions in the
  purchase of goods and services, listing only those with an
  individual share $>2$\% (which together account for more than 95\%
  of the total). The first column lists the identifier in the
  French NACRES nomenclature.}
\label{tab:purchase} 
\end{table*}

\begin{figure}[!t]
\begin{center}
\includegraphics[width=0.7\textwidth]{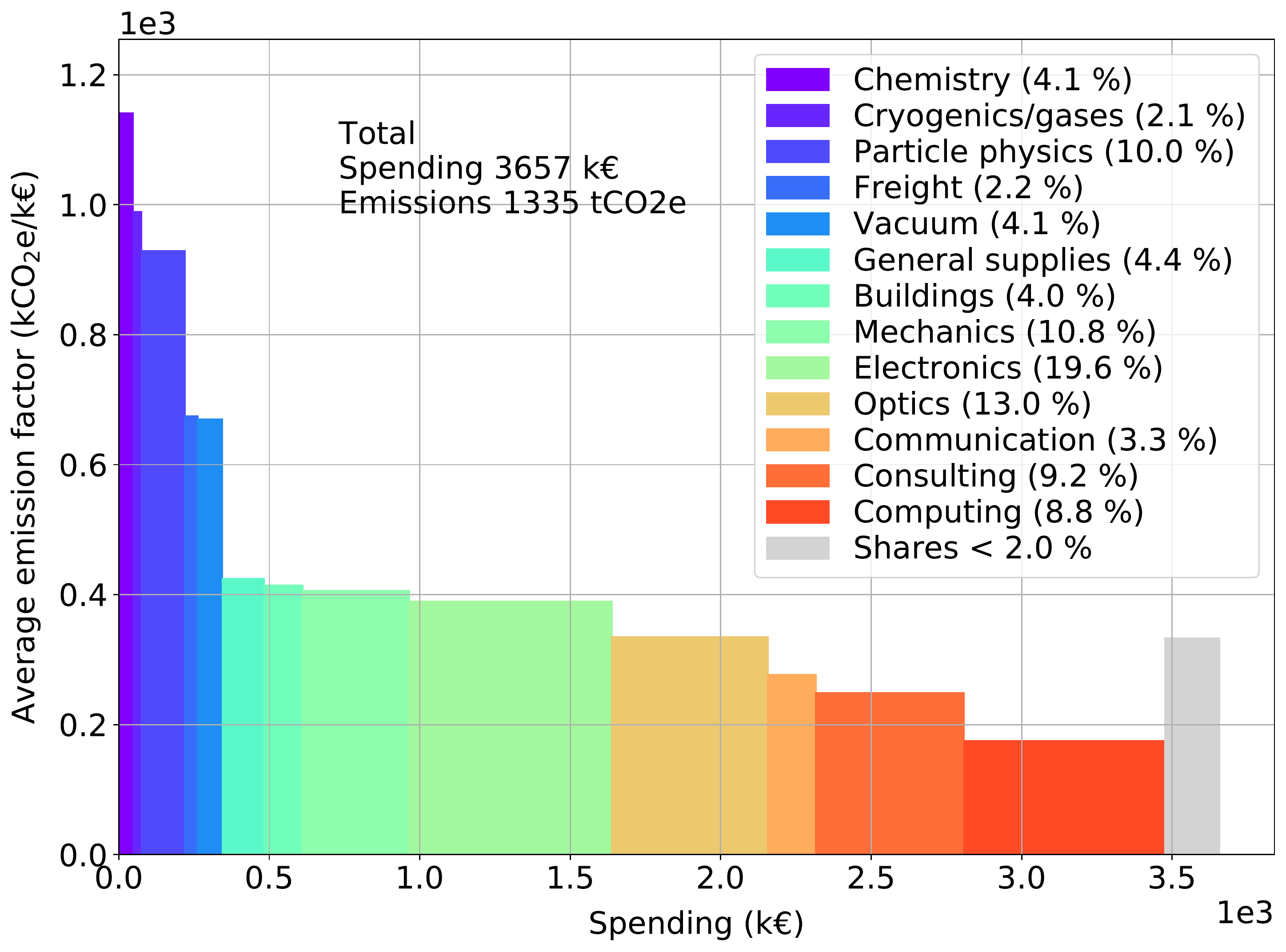} 
\caption{Average emission factors for the purchase of goods and
  services, organized by the main categories of expenditure, as a
  function of the cumulative expenditure. The size of the rectangles is proportional to the GHG emissions. The last gray block
  aggregates all shares that individually contribute less than 2\% of
  the total GHG emissions. The fractional contribution to the GHG
  emissions for each category is shown in parentheses.}
\label{fig:purchase} 
\end{center}
\end{figure}

\noindent As mentioned in Sect.~\ref{method:carto}, our activities
rely on important input flows of material and services,
e.g. the purchase of electronic equipment, and subcontracting the design of
certain instrument components. Not included in this category are the
acquisition of computers, use of external numerical resources,
personal travel services, and use of observational data since we deal
with these sources using specific approaches in other sections below.\\

\noindent Ideally, the environmental impact of goods and services
purchased in the context of our activities would be assessed and
communicated by the suppliers. This so-called supplier-specific
information, however, rarely exists and it is beyond the scope of our
work to calculate it {\em ab initio} since it would require conducting
complete GHG accounting for a wide diversity of products, based on the
physical flows needed to produce them. Instead, we used a cost-based
approach that converts the economic value of the purchased goods and
services into an estimate for the associated GHG emissions using
economic sector-average emission factors (typically equating a
k\euro\ or M\euro\ of expenditure on a given family of products,
e.g. electronic equipment or software development, to \kco or
\tco).\\

\noindent In practice, our activity data for this category are IRAP's
annual expenditure through the Paul Sabatier University or CNRS,
sorted into about 2000 categories that are labelled according to a
specific code (NACRES, standing for Nomenclature Achats Recherche
Enseignement Sup\'erieur). We associated these expenditure categories
with emission factors obtained from ADEME for about 35 broad activity
classes \cite{ademe-ratios}. We tested an alternative set of $\sim400$
factors from the United States Environmental Protection Agency (EPA)
\cite{epa:2020}, with a much finer description of industrial and
economic activities. The difference between the final results
calculated using the ADEME and EPA factors was at the 25\% level, and
there was overall good agreement in the distribution of emissions over
the different expenses\footnote{The EPA set of emission factors for
  the different GHG gases was converted into \coe using
  global warming potential values for a 100-year time horizon from the
  GHG Protocol \cite{gwp:2016} and the spendings in 2018 US dollars
  were translated into 2019 euros using a currency exchange rate of
  0.911.}. We therefore kept the ADEME set of emission factors since
it is more relevant to France than the EPA set. The expenditure-based
approach is admittedly the least precise approach in GHG accounting,
but it offers a way forward when other GHG accounting techniques cannot be
achieved within a reasonable amount of time.\\

\noindent The expenditure on goods and services at IRAP for 2019
amounts to 3657\,k\euro, excluding all expenses that are dealt with in
a more specific way in other sections of this paper (e.g. energy, computers,
travel etc). The uncertainty on total expenses is assumed to be zero,
and the final uncertainty on GHG emissions is computed as the
quadratic sum of uncertainties over all broad purchase categories,
assuming for the latter a relative uncertainty of 80\% as recommended
by ADEME and the \gls{bc} method. Such an approach in computing the
uncertainty is justified by the large number of different purchases,
with a mix of tailor-made products and mass-consumption items.\\

\noindent The emissions from the purchase of goods and services amount
to $1335 \pm 342$\tco. These GHG emissions divided by the expenditure
correspond to an average emission factor of 365\kcoperkeuro, which is
of the same order as the emission factor for electronic, optical, and
computer equipment in the ADEME database ($\sim400$\kcoperkeuro). The
distribution of the emissions into broad purchase categories is
presented in Table \ref{tab:purchase}, and the corresponding average
emission factors are illustrated in Fig. \ref{fig:purchase}. The bulk
of the emissions is due to the purchase of technical
equipment, most likely in connection to the numerous instrument
development projects based at IRAP. The share related to the local
infrastructure amounts to $\lesssim10-15$\% (obtained by summing
categories A, B, and a fraction of E).\\

% Computer equipment
\subsection{Computer equipment}
\label{res:computer}

\noindent We treat the purchase of computer equipment separately from
the purchase of goods because it allows a finer and more reliable
assessment of the associated GHG emissions. The formal approach to
this contribution is to estimate the total equipment in use at the
institute during our reference year 2019, and to add up the GHG
emissions associated with each machine for an amortization period of
typically 4-5 years (which means that machines older than that are not
counted, while others have a yearly contribution corresponding to the
total GHG emissions due to their production divided by the
amortization period). In practice, it was impossible to implement such
a calculation owing to the difficulty that we encountered to obtain a
robust census of all our computer equipment. Instead, we summed the
number of computers and other main peripherals purchased over the period
2018-2020, and computed yearly GHG emissions per type of equipment
assuming an amortization period of three years. Such an extraction was
made by an automated scanning of purchase listings, but may not be
highly accurate because of the lack of uniform, clear labels in the
purchase orders. From a visual inspection of the labels and expenses
of the different purchases, we consider that the volume of equipment
eventually counted in our report is a lower limit, which probably
underestimates the true value by at most 10\%.\\

\noindent The emission factors for different computer equipment and
peripherals were taken from the ECODIAG tool provided by the ECOINFO
dedicated working group of the CNRS \cite{ecodiag:2022}. They
correspond to cradle-to-gate emissions, and do not include end-of-life
impacts, while power usage during the equipment lifetime is counted in
our electricity consumption (except for the small part used when the
equipment is taken out of the institute). ECODIAG offers emission
factors for a variety of models for each type of equipment and we
selected those labeled as default, which seem to correspond to the
typical equipments mainly in use in the French astrophysics research
community. Nevertheless, we also extracted a range of emission factors
that characterise the diversity of equipment in use at the institute,
to illustrate the possible uncertainty range due to the imprecise
identification of each item. Emission factor variations depending of
the equipment model are typically of the order of 20-40\%, or higher
in the case of PC/workstations. In addition, ECOINFO warns that they
can increase up to threefold depending on the set of options chosen
for the equipment. We eventually retain a typical uncertainty on
emission factors of 50\%, following the recommendation by ADEME.\\

\noindent The results are presented in Table \ref{tab:computers}. We
purchase every year about 120 computers (at least over 2018-2020), for
a staff of about 260 people, which suggests a typical lifetime of
about 2-3 years per computer and/or a progressive inflation of the
average number of computer equipment per person if the actual lifetime
is longer. This leads to GHG emissions of $81 \pm 40$\tco.\\

\begin{table*}[t!]
\centering
\begin{tabular}{ccccc}
\hline
Equipment & Number & Factor & Range & Emission \\
 &  &  (\kco/unit) & (\kco/unit) & (\tco) \\
\hline
Laptop & 71 & 300 & 250-370 & 21.3 \\
PC/Workstation & 45 & 600 & 170-650 & 27.0 \\
Server & 6 & 1300 & - & 7.8 \\
Standalone screen & 10 & 430 & 350-590 & 4.3 \\
Tablet & 7 & 150 & - & 1.0 \\
\hline
Total &  &  & & 80.8 \\
\hline
\end{tabular}
\caption{Distribution of GHG emissions associated with different types
  of computer equipment and peripherals. The numbers of items
  correspond to the average over the period 2018-2020.  Total
  emissions include one screen per workstation, on top of the
  standalone ones. No emission factor range is given for servers and
  tablets on the ECODIAG tool.}
\label{tab:computers} 
\end{table*}

% External computing, storage, and data flow
\subsection{External computing, storage, and data flow}
\label{res:num}

\noindent We estimated the usage of external resources for computing
and data storage via our online survey (Sect.~\ref{res:survey}). For
computing, we estimate a total usage of 7~MhCPU. We convert this into
GHG emissions using the emission factor of 4.68~\gco/hCPU estimated by
\cite{berthoud:hal-02549565} for a computing center in Grenoble with
an uncertainty of $80\%$. It is appropriate to use an emission factor
based on a French computing center (low carbon impact of electricity)
because $>99\%$ of the computing occurs in centers located in
France. This yields $\sim$33~\tco of GHG emissions.\\

\noindent We took a similar approach for external data storage. In
this case, we estimate a total of $\sim$39~TB of data stored in France
and $\sim$254~TB of data stored in other countries. The carbon
footprint of data storage is highly uncertain and depends on the type
of storage (short-term or long-term), the occupancy rate of the
storage centers, and the carbon intensity of the electricity. We adopt
the overall emission factors estimated in Ref. \cite{berthoud2021}. For
France we have an interval of 7 to 40~\gco/GB/yr and we adopt a
representative value of 25~\gco/GB/yr. Based on their study of the
impact of the electricity carbon intensity, we assign to data storage
in countries other than France (mostly U.S.A.) a carbon footprint
which is four times larger, namely an interval of 28 to 160~\gco/GB/yr
with a representative value of 100~\gco/GB/yr. This yields a
representative value of 26~\tco for the associated GHG emissions,
within an interval of 4 to 63~\tco.\\

\noindent Last, we estimated the impact of data transfer over the
global network, generated by a variety of day-to-day use of numerical
services such as exchanging emails, videoconferencing, or downloading
large astronomical data sets. Monitoring of the ingoing and outgoing
data flows at IRAP is not performed with a sufficient accuracy, so we
had to assume a typical amount of data transferred over the network
over a year. As a representative number, we used the total amount of
data stored externally, 293\,TB ($\pm 50\%$), because these data had
to be transferred at least once and such a typical volume of about
1TB/capita/year encompasses the data flows generated by the regular
use of the network by one individual over a year: thousands of emails
(hundreds of which with MB-sized attachments), hundreds of hours of
audioconferencing or videoconferencing (with data transfer rate of
30-100\,MB/h and 500-2400\,MB/h, respectively), dozens of hours of web
browsing (with data transfer rate of 50-300\,MB/h), and downloads of
all kinds (papers, software installations or updates, astronomical
datasets). There is most likely a large, order-of-magnitude, scatter
in data transfer from one individual to another depending on staff
category (e.g. in time spent in videoconferencing or amount of data
downloaded) and usage profile (e.g. audioconferencing versus
high-definition videoconferencing). Nevertheless, our purpose here is
to estimate the typical weight that data transfer represents in our
final footprint, and the results can easily be rescaled to an actual
use case. Ultimately we find that data transfer represents only a very
small contribution to our carbon footprint. \\

\noindent The amount of GHG emissions is obtained by associating the
volume of data transferred to an electricity consumption, and then
with an average carbon intensity for electricity. We considered an
emission factor combining both and obtained in Ref. \cite{Renater:2021},
about 1-2\gcopergb, for the transfer of data between Orsay and
Montpellier in France via the RENATER network. The estimate includes
the impact of power consumption, manufacturing and installation of the
equipments, and network supervision activities. The cost of data
transfer increases strongly with distance/number of nodes between
emitter and receiver, and carbon intensity of electricity. Conversely,
it decreases with increasing load of the network and increasing
lifetime of the equipments. To be more representative of international
data transfer, over long distances and consuming electricity that has
on average a much higher carbon intensity than in France, we
considered a typical range of 2-10\gcopergb as emission factor, with a
baseline value of 5\gcopergb. The latter value is consistent with that
extrapolated for 2019 in Ref. \cite{Aslan:2017}. Eventually, this
yields yearly GHG emissions of about 1.5\tco for all IRAP staff, with
a likely range 0.3-3.2\tco.

% Research infrastructures
\subsection{Research infrastructures}
\label{res:infra}

\noindent Scientists from our institute use a long list of research
infrastructures, such as space telescopes, space probes and
ground-based observatories. According to the \gls{bc} method, the
carbon footprint associated with the use of these facilities needs to
be included in the institute's carbon footprint estimate. Details of
the carbon footprint estimation method for research infrastructures
are presented separately in Ref. \cite{Knodlseder:2022}. In short, we
identified the facilities that were used from all refereed
publications co-authored by IRAP scientists that were published in
2019, resulting in a list of 46 space missions and 39 ground-based
observatories. We estimated the carbon footprint of each facility
using monetary ratios, and consolidated the results for space missions
with an alternative estimate based on the payload wet mass. Dedicated
emission factors for this analysis were derived from published carbon
footprint assessment reports. For space missions, we derived an
emission factor of 140~\tcopermeuro mission cost and
50~t~CO$_{2\textrm{eq}}$/kg payload wet mass, for ground-based
observatories we derived 240~\tcopermeuro construction cost and
250~\tcopermeuro operating costs. We note that these emission factors
are on the low side of the sector-based estimates provided by ADEME,
hence it seems unlikely that our adopted values are significantly
overestimated.\\

\noindent As activity data, we gathered full mission cost and payload
launch mass estimates for space missions and construction, and yearly
operating costs for ground-based observatories from publicly available
documents. These data were complemented by information provided
directly by some of the facilities, and a parametric cost model for
some of the 1--2 metre class telescopes in our list. All activity data
collected are provided in the supplementary information of Ref.
\cite{Knodlseder:2022}. Based on these activity data and the
aforementioned emission factors, we computed lifecycle and annual carbon footprint
estimates.\\

\noindent To determine which fraction of the research infrastructure
carbon footprint should be attributed to our laboratory, we use two
methods. In the first method, we multiply the annual carbon footprint
of an infrastructure with the fraction of peer-reviewed papers in 2019
that have authors affiliated to IRAP. We determined this fraction from
the Astrophysics Data System (ADS) using a full text search for the
year 2019. For this purpose, we constructed a dedicated query string
for each infrastructure with the aim to cover as many
infrastructure-related publications as possible while keeping the
false positives at a minimum.  This results in a footprint of $20\pm3$
\ktco for IRAP in 2019. Attributing this footprint equally to the 144
astronomers with PhD degree that worked at IRAP in 2019 results in
$139\pm23$ \tco per astronomer. If we instead divide the annual
research infrastructure carbon footprint equally by the total number
of staff that worked at IRAP in 2019 (263 people), we obtain $76\pm12$
\tco per IRAP staff member.\\

\noindent We note that this attribution method does not provide IRAP's
share of the total carbon footprint of research infrastructures among
all existing astronomical institutes in the world. Since scientific
articles are often signed by authors from multiple institutes, each of
these institutes will get the same attribution, implying that the sum
of all attributions will exceed the total carbon footprint of all
research infrastructures. The share can however be estimated by
replacing the number of peer-reviewed papers by the number of unique
authors, i.e. multiply the annual carbon footprint of an
infrastructure with the fraction of unique authors of peer-reviewed
papers in 2019 that are affiliated to IRAP. This results in a carbon
footprint of $4.0\pm0.7$ \ktco for IRAP in 2019. For each of the IRAP
astronomers with PhD degree this corresponds to a footprint of
$27.4\pm4.8$ \tco, for each person working at IRAP in 2019, the
footprint is $15.0\pm2.6$ \tco. We note that some double-counting may
still occur using this approach since an individual may be affiliated
to multiple institutes. But for a given individual, the computation of
the share should be accurate.

\subsection{Emission sources excluded from our scope}
\label{res:out}

\noindent Some sources of GHG emissions were not included in our assessment, or
only partially, or not in the way they should have been according to
the principles of the \gls{bc} method. The main reason in most cases
was the unavailability of, or difficulty in obtaining access to, the relevant
data. We list these sources here to provide a complete picture of what
should be the full scope of the assessment, and remind that the final
carbon footprint we report here is formally a lower limit. All the items
listed below constitute possible avenues for improvement of future
assessments.\\

\textbf{People}: This should include people attending events organized
by the institute, such as workshops or conferences. Doing so would
likely increase the institute's GHG emission, but it would allow
the institute to identify a potentially powerful lever arm. IRAP, as
the sole or primary organizer of an event, could opt for relocating it
to a location that minimizes the total travelled distance, replacing
it by a virtual gathering, or, for a recurrent event, reducing its
frequency. All such measures would have a very efficient overall
return. In practice, however, no major conference or workshop was
included in the present assessment because: (i) IRAP was, to our
knowledge, rarely the sole or primary organizer of such an event in
2019; (ii) collecting the corresponding travel data for all
participants was unfeasible. Ultimately, the only cases we took into
account were trips by invited researchers (e.g.  seminar speakers)
because their travel and accommodation was paid for by the institute.

\textbf{Material}: According to the \gls{bc} methodology, delivery and
freight are supposed to be handled as a distinct category, to separate
the impact of producing goods from that of transporting them. The
expected quantity here is typically the amount of mass transported
over a total distance, split into transportation means (rail, road,
air, sea). In practice, it was impossible to get such detailed
information for all incoming material (e.g. electronic or optical
components) and outgoing equipment (e.g. delivery of instruments or
parts of instruments). A part of the freight is indirectly included
via the purchase of transportation services, but with a monetary
approach, rather than a physical one.

\textbf{Services}: It is recommended to include the GHG emissions
associated with the use of services provided by IRAP to the outside
world, e.g. open databases and public software developed and/or
maintained at IRAP. The rationale for including this contribution is
the same as for conference organization : reducing the subsequent
emissions by optimization at the source as much as possible. For
numerical services, we do not keep an exhaustive census of the
software and databases provided by IRAP, with homogenous statistics
about their use. Partial information exists in the case of some
community services supported by the institution (the Services
Nationaux d'Observations). Assessing the impact of the data storage
and transfer generated by IRAP's SNO would be an interesting
avenue to explore in future assessments.

\textbf{Internal support}: IRAP relies on support activities, such as
administration, maintenance, and financial and insurance services
provided by higher-level institutions (mainly the Paul
Sabatier University and the CNRS). Not all of these activities are local and easily
integrated into our assessment. The \gls{bc} methodology also requires
to include the carbon impact of the construction of buildings, with a
specific treatment for amortization over typically 10-20
years. According to this rule, most of our buildings would already be
amortized, but some recent installations should have been taken into
account but were not. We can however provide an order of magnitude for
the most recent building construction in our institute, the
``Plateforme d'Ing\'enierie et d'Instrumentation Spatiale (P2IS)'',
inaugurated in 2015. It has a floor surface of 368\,m$^2$ and consists
of a mix of offices and technical rooms. According to the ADEME
database, emissions from construction range from 650\kgcopermsq for
office buildings to 825\kgcopermsq for industrial buildings. This
implies total GHG emissions in the range $239-304$\tco for the
construction of P2IS, or about $12-15$\tcoperyr assuming an
amortization period of 20 years, which is a relatively minor
contribution in our total footprint. This assessment should however be
performed for all extraordinary operation or acquisition, e.g. the
purchase of a large equipment for instrumental development. In the
absence of a well-established inventory and an easy way to associate
these to GHG emissions, we excluded them from the scope of this
assessment.

% Synthesis
\section{Synthesis}
\label{synth}

% The carbon footprint of IRAP
\subsection{The carbon footprint of IRAP}
\label{synth:irap}

\begin{figure}[!t]
\begin{center}
\includegraphics[width=0.85\textwidth]{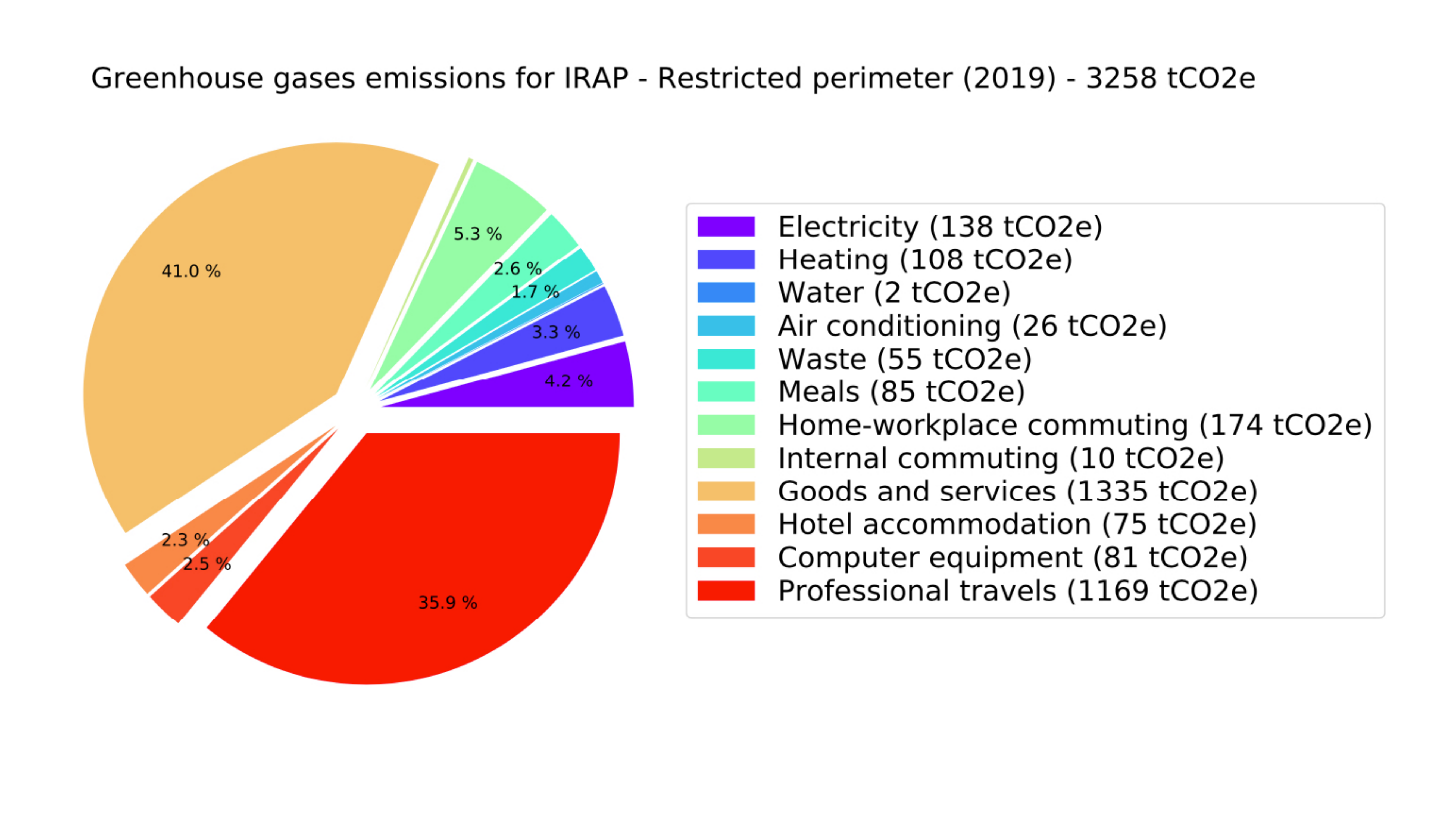}
\includegraphics[width=0.85\textwidth]{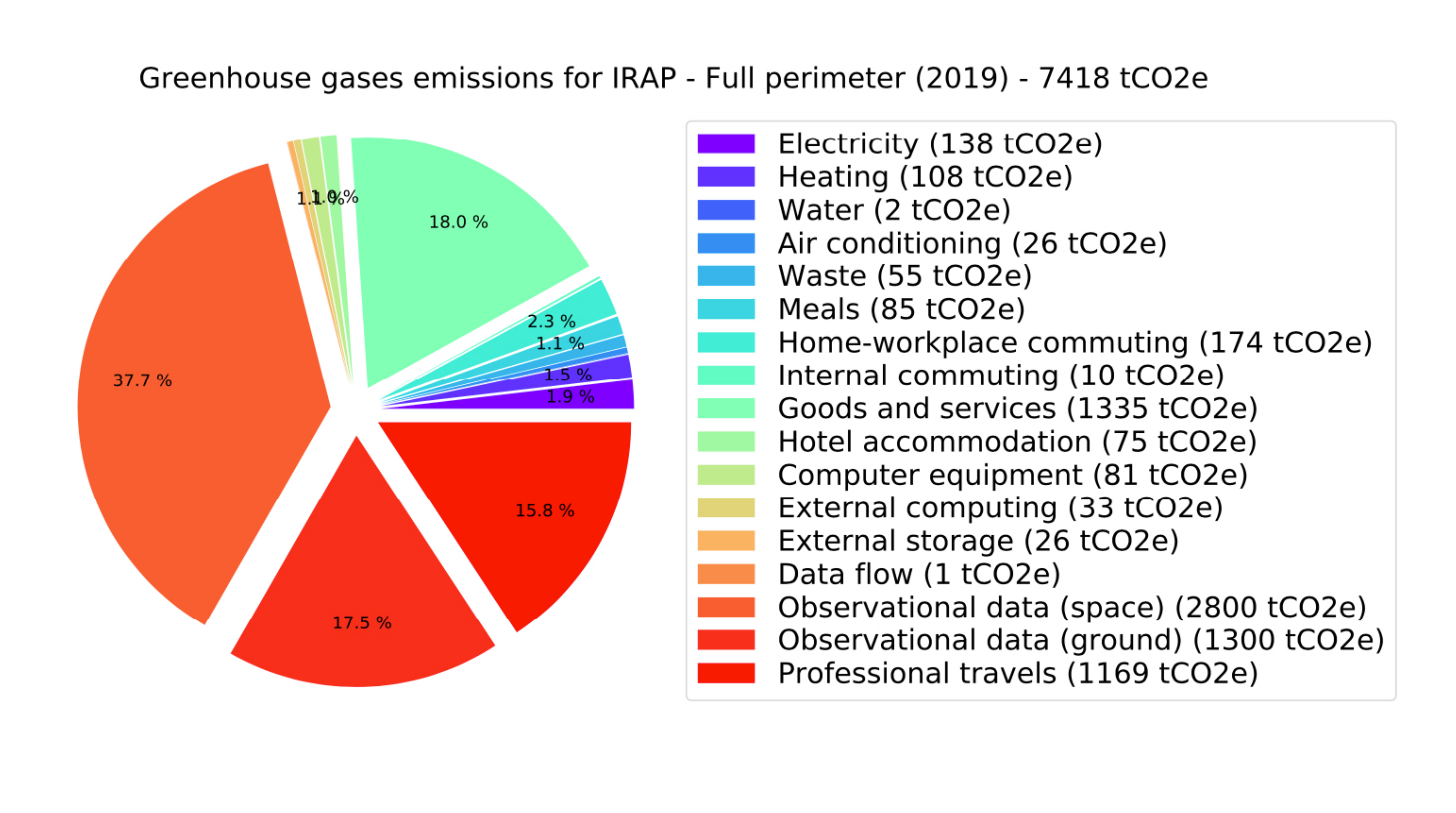}
\caption{Distribution of IRAP's GHG emissions in 2019, for the
  restricted and full perimeter, without and with the use of external research
  infrastructures, respectively (top and bottom). Only wedges with a
  relative contribution above 1\% are labelled.}
\label{fig:pie}
\end{center}
\end{figure}

\noindent The breakdown of all IRAP's GHG emissions per source is
summarized in Table~\ref{tab:summary} and presented in
Fig.~\ref{fig:pie}. The final estimated quantity for 2019 over the
full perimeter of the assessment is about 7400\tco, or an average of
28\tcoperyr per staff member. We also defined a restricted perimeter,
by leaving out the contribution of external research infrastructures
including external computing and storage. Over this restricted
perimeter, our emissions amount to 3300\tco or an average of
13\tcoperyr per staff member. The rationale for this distinction is
that IRAP has significant operational control over the restricted
perimeter, but a more limited one over external research
infrastructures, so such a separation is relevant in view of a local
reduction plan. This practical distinction is however relevant only to
the means and strategies for implementing reduction measures, and
should not elide that IRAP's total 2019 footprint is 7400\tco.\\

\noindent The operation of the local infrastructure -- heating,
electricity, commuting, food, waste, and $\sim10-15$\% of the purchase
of goods and services -- makes a relatively small contribution to
IRAP's carbon footprint, about 800\tcoperyr. Thus, while it remains
relevant to reduce IRAP's consumption of electricity and gas
consumption and to switch away from conventional cars for local
transport, initiatives focussed on the institute's local
infrastructure will not be sufficient to achieve emission reductions
that are compatible with France's 2050 targets (i.e. reducing
emissions by a factor of 5). We note that the relatively small
contribution of local infrastructure may be specific to IRAP, and that
other astronomy institutes may have quite different emission
profiles. The energy used to produce the institute's electricity and
heating has a relatively low carbon footprint, with electricity being
predominantly of nuclear origin in France and heating of our largest
building arising from biomass burning\footnote{For more information
  about the use of biomass as an energy source, and its actual carbon
  footprint and possible role in reduction strategies, see
  \url{https://easac.eu/publications/details/key-messages-from-european-science-academies-for-unfccc-cop26-and-cbd-cop15/}},
corresponding to emission factors of 0.06-0.07\kgcoperkwh. Assuming a
worst-case carbon intensity of $\sim0.8$\kgcoperkwh instead, typical
of Australia, the related sources would amount to about 2700\tcoperyr,
comparable to the sum of professional travels and purchase of goods
and services at IRAP. A higher carbon intensity of electricity would also
affect other sources such as external computing, and to some extent
the purchase of goods and services. If IRAP were situated in a country that
relies significantly on fossil fuels for electricity production and
heating, we estimate that our total 2019 footprint
would be at least 10000\tco, or 38\tcoperyr per staff member.\\

\noindent The most significant sources of GHG emissions at IRAP,
ordered by decreasing contribution, are: the use of astronomical
research infrastructures, the purchase of goods and services, and
professional trips, with a predominant share of air travel in the
latter. As mentioned above, the bulk of the emissions from the
purchase of goods and services is attributable to the numerous
instrument development projects performed at IRAP. Together with the
preponderant share from the use of observational data, this clearly
points to the main driver of our large carbon footprint: astronomical
research infrastructures. More precisely, it is the current practices
related to conceiving, designing and developing astronomical research
infrastructures, and the scale and cadence at which we deploy
them. Moreover, it is likely that a significant fraction of
professional trips at IRAP are also connected to instrument
development projects, which makes the conclusion even stronger
(although we could not check this point owing to the lack of the
relevant information).\\

\noindent Whether such a trend can be considered as generic in the
community remains to be confirmed by performing exhaustive carbon
footprint assessments at other institutes. IRAP has a long history in
instrument development and observational data analysis, which
certainly weighs in the present result. Institutes that specialise in
theory and numerical simulations may have a smaller footprints,
although this will depend on the amount of supercomputing performed
and the carbon intensity of the electricity powering the computing
clusters. Ultimately, however, such a distinction may be irrelevant
since research in astronomy and astrophysics requires all of these
activities (theory, simulations, observations). What is needed is a
community-scale strategy for our use of all resources and facilities.

% Comparison and discussion
\subsection{Comparison and discussion}
\label{synth:compa}

\noindent IRAP's total emissions in 2019 ($\sim$7400\tco) correspond
to 28\tcoperyr per staff member, on average, from professional
activities only. This can be compared to the average footprint of a
French resident (combining private and professional emission sources)
of 11\tcoperyr, although that comparison is not directly meaningful as we will discuss below.\\

\noindent More relevant is a comparison with other
institutes. Ref. \cite{Jahnke:2020} estimated that the annual emissions of
the Max Planck Institute for Astrophysics (MPIA) located in
Heidelberg, Germany, amounts to 18\tcoperyr/researcher, and
$\sim9$\tcoperyr/staff member if the load is shared over all
employees, including all support staff. Therefore, work-related emissions only are 60\%
higher than the average for German residents, 12\tcoperyr, which
includes the impact of both professional and private activities. Nearly half of the MPIA
emissions arise from air travel, and a quarter from the local
electricity consumption of supercomputing
\cite{Jahnke:2020}. The latter point is an interesting difference between MPIA and IRAP: the emissions from supercomputing at IRAP amount to 0.2\tcoperyr per researcher, on average, and this covers the impact from electricity consumption, equipment, and operations of the computing centers; at MPIA, supercomputing generates an average 4.6\tcoperyr per researcher, from electricity consumption only. Only a small part of the difference can be explained by the carbon intensity of electricity, which is a factor $\sim 4$ higher at MPIA with respect to IRAP.
\\

\noindent The situation of Australian astronomers is more alarming, with emissions
amounting to 42\tcoperyr/researcher, or twice the average impact of an
Australian resident, 21\tcoperyr \cite{Stevens:2020}, which already
ranks among the highest on the planet. We emphasise, however, that the
perimeter adopted in the assessments by \cite{Stevens:2020} and
\cite{Jahnke:2020} is much more restricted than the one used here.\\

\noindent To put these numbers in a global perspective, the world
average rate of GHG emissions today is $\sim5-6$\tcoperyr per capita. To
achieve carbon neutrality, i.e. for anthropogenic emissions to be
absorbed by natural carbon sinks from land and oceans, the global
average must be reduced to 2\tcoperyr per capita by 2050. The average
emission rates compatible with carbon neutrality will be even lower if
the planet's capacity to capture carbon dioxide has been strongly
degraded by 2050 and/or if the world population has inflated strongly
beyond 10 billion inhabitants \cite{wpp:2019}.\\

\noindent There is ongoing debate in the community about the exact
meaning of these emission rates per capita, and the way that the
burden of GHG emission reductions should be shared. A common response
from those interested in maintaining the {\it status quo} is to argue
that the GHG emissions from astronomy research should be attributed
across a much larger population base than researchers and the staff of
astronomical research institutes, i.e. over the full production chain
including the suppliers of goods and services, or over the general
population since the benefits of astronomy research are enjoyed by
society at large. However, we stress that in carbon accounting, the
objective is not to uniquely identify emissions with a given group of
individuals, but rather to quantify all emissions that an activity
depends upon to exist and generates while it is in operation. The
ultimate goal is to identify all possible avenues for GHG emission reduction. \\

\noindent Our assessment shows that performing research in astronomy
and astrophysics at IRAP in 2019 stimulated GHG emissions equivalent
to 28\tcoperyr per person involved in that activity, on average. That
these emissions are spread across a variety of social and economic
sectors is obvious, and should not be used as an argument to lower our
impact by distributing it over a larger population base. 
One way of viewing the global average target of 2\tcoperyr per capita
by 2050 is as an average budget that should not be exceeded, and
within which societies should fit what they deem necessary to human
life. In a democratic debate about how this target should be achieved
and which activity sectors should be afforded permission to exceed the
average value, the place of research should be discussed (alongside
all other sectors) with a quantitative estimate of their costs and
benefits. Here we estimate that the carbon cost of astronomy research
is 28\tcoperyr per person employed in that activity when it is
performed according to IRAP's 2019 standards.
Obviously, not all activity sectors in society (including research) can be accorded a footprint that exceeds the average allocated budget,
otherwise the latter will rapidly be exceeded. Shifting our professional practices to bring the average carbon
footprint of astronomy research closer to the target would thus seem an important step
to guarantee the future of our field.

% The way forward
\section{The way forward}
\label{discu}

% Easing carbon accounting
\subsection{Facilitating carbon accounting}
\label{discu:carbon}

\noindent Regardless of the strategies that are adopted to reduce our
emissions, regular assessments of the institute's carbon footprint to
take stock of the situation, monitor progress, and adjust the
emissions reduction plan will be required. During our work to estimate
the carbon footprint of IRAP, it became clear that a key challenge
lies in the availability of activity data. The problems we faced
ranged from information being totally absent (e.g. fine-grained
measurements of the numerical data flows generated by our activity),
to data being incomplete (e.g. professional travel data missing the
departure/arrival locations or the reasons for travelling), or tedious
to collect (e.g. constructing a list of computer equipment
purchases by hand). In this section, we present some recommendations for how
to improve this situation and move towards seamless -- or at least
less painful -- carbon accounting. These recommendations may not be
directly relevant for astronomy institutes based outside of France.\\

\noindent Preparing the information in a complete and relevant format
at the source is an urgent short term goal. National suppliers
responsible for providing services for travel and hotel reservations,
computer equipment and catering are ideally positioned to amalgamate
global information that can be extracted for later use at different
levels, from small entities such as institutes, to research
federations such as the ``observatoires des sciences de l'univers'',
and umbrella institutions like the CNRS. Such a requirement should be
rapidly negotiated at the national level.\\

\noindent At the level of campuses, information pertaining to
infrastructure such as electricity, heating, water and waste should
also be collected centrally and shared among all users of the same
site. When a specific mode is used to supply heating, the relevant
emission factor should be updated according to the actual operation of
the facility (e.g. actual mix of gas and wood in the case of our
university).\\

\noindent At the institute level, it is desirable to have activity
data organized in the most appropriate way to raise awareness and
trigger action. In a laboratory like IRAP, knowing the carbon
footprint of different IRAP-based projects would help staff members
appreciate their actual contribution to the institute's emissions, as
well as the required effort to be made. Some development of the software used
by our administration is needed to easily associate activity data such as
travels and purchases to the relevant department or project within the institute,
all this while preserving anonymity of the individuals.\\

\noindent Facilitating carbon accounting requires information that has
not previously been collected, or revising the format of the collected
information.  At least initially, this will imply an increased
administrative burden. Care should be taken to not overload our
colleagues and ensure that enough time is allocated for them to
contribute to this high-priority duty.\\

\noindent The source of GHG emissions at IRAP with the largest
uncertainties, both in terms of activity data and emission factors,
was activity related to digital technologies. Our estimates of the use
of external storage and data transfer are highly uncertain, as are the
estimates for their emission factors, with order-of-magnitude scatter
frequently encountered in the literature. Although these activities
made a relatively small contribution to our footprint (in part because
of the low carbon intensity of electricity in France), the rapidly
evolving nature of this field, and in particular the strong and
continuous increase in many practices, e.g. the volume of data stored
in clouds, massive computing, data flows, requires a careful
assessment of its impact, to make sure that it remains under control
in terms of environmental sustainability. On the side of activity
data, more measurements by the Information \& Technology services are
clearly needed for us to have a better understanding of the actual
situation. An assessment of the ingoing and outgoing flows of data,
split into usage, is clearly missing today to guide us towards a
better practice. On the side of emission factors, expert work is
required to fully appraise the relevant emission factors appropriate
at a given time, location, and for a given usage. In France, such an
effort is conducted by the ECOINFO research group, and some of their
pioneering studies were critical to our estimate. We recommend strong
and lasting support from our institutions to this group and, going
beyond carbon accounting, regular staff training about the
environmental impact of digital technologies.\\

\noindent Finally, as already emphasised and recommended in Ref.
\cite{Knodlseder:2022}, there is a crucial lack of reliable
information about the carbon footprint of ground-based observatories and
space-borne instruments. The first estimates provided in that paper
and performed in the context of the present assessment for IRAP
suggest that it is a large, if not dominant, share for most
institutes. We therefore encourage the entities running these research
infrastructures to rapidly assess their total footprint, including
construction and operation, and to publicly share them so that all
institutes can include the information in a uniform way in their carbon
footprint assessments.\\

\noindent We emphasise that widespread carbon accounting by all actors
involved in astronomy research is just a step towards solutions, and
not the solution itself. There is a growing risk that carbon footprint
assessment becomes essentially a communication measure, while the hard
decisions that must be made to reduce emissions are left
aside. Transitioning from fair and relevant carbon accounting to
far-reaching reduction strategies requires institutional mechanisms
that are yet to be envisaged or implemented. In that respect, the situation
in France is enlightening: carbon accounting has been mandatory for
more than a decade and companies and institutions of all kinds have
published their carbon footprint, but this requirement appears to have
had zero impact on France's GHG emission curves. 
The reasons are manifold, but the conclusion is inevitable : 
carbon assessments without emission reduction strategies are ineffective.

% Possible avenues for reduction
\subsection{Possible avenues for reduction}
\label{discu:reduc}

\begin{table*}[t!]
\centering
\begin{tabular}{c c  c  }
\hline
Scenario & Emissions & Gain \\
 & (\tco) & Rel. Restr. Full \\
\hline
\multicolumn{3}{c}{Professional travels} \\
\multicolumn{3}{c}{(1169\tco, 35\% restricted total, 16\% full total)} \\
\hline
1 - Train in FR & 952 & -19\% / -6.7\% / -2.9\% \\ 
2 - Plane 2 non-EU & 977 & -16\% / -5.9\% / -2.6\% \\ 
3 - Train in FR, Plane 4 EU+2 non-EU & 719 & -38\% / -14\% / -6.1\% \\ 
4 - Train in FR, Plane 2 EU+1 non-EU & 508 & -57\% / -20\% / -8.9\% \\ 
\hline
\multicolumn{3}{c}{Commuting} \\
\multicolumn{3}{c}{(184\tco, 5.7\% restricted total, 2.5\% full total)} \\
\hline
1 - $<$2.5km bike/foot, for the rest -20\% car & 155 & -16\% / -0.9\% / -0.4\% \\ 
2 - $<$5km bike/foot, for the rest -40\% car  & 125 & -32\% / -1.8\% / -0.8\% \\ 
3 - same as 2 with 50\% electric/hybrid cars & 103 & -44\% / -2.5\% / -1.1\% \\ 
4 - 3 days remote working & 152 & -17\% / -1.0\% / -0.4\% \\ 
\hline
\multicolumn{3}{c}{Meals} \\
\multicolumn{3}{c}{(85\tco, 2.6\% restricted total, 1.2\% full total))} \\
\hline
1 - 50\% standard meals $\mapsto$ flexitarian & 65 & -24\% / -0.6\% / -0.3\% \\ 
2 - standard meals $\mapsto$ flexitarian/vegetarian & 37 & -56\% / -1.5\% / -0.7\% \\ 
3 - 100\% vegetarian meals & 22 & -74\% / -1.9\% / -0.9\% \\ 
\hline
\multicolumn{3}{c}{Computer equipment} \\
\multicolumn{3}{c}{(81\tco, 2.5\% restricted total, 1.1\% full total))} \\
\hline
1 - 4-year lifetime for computers & 59 & -27\% / -0.7\% / -0.3\% \\ 
2 - 6-year lifetime for computers & 50 & -38\% / -1.0\% / -0.4\% \\ 
\hline
\multicolumn{3}{c}{Heating and electricity} \\
\multicolumn{3}{c}{(246\tco, 7.6\% restricted total, 3.3\% full total)} \\
\hline
1 - SNBC national strategy & 125 & -49\% / -3.7\% / -1.6\% \\ 
\hline
\end{tabular}
\caption{Benefits from various reduction scenarios over a selected set
  of GHG sources. The gain column lists the gain for
  the source of emission under consideration, then the gain relative
  to the restricted perimeter, and lastly the gain relative to the full
  perimeter.}
\label{tab:reduction} 
\end{table*}

\noindent At the moment of writing, there is no clear target that has
been defined for the reduction of emissions from scientific research,
or by astronomy in particular. To gauge the magnitude of the necessary
reductions, we can consider France's national reduction targets of
50\% by 2030 and 80\% by 2050. However, these figures do not account
for the fact that some activity sectors of our society will likely be
allowed to continue producing a higher level of GHG emissions, and
that other activity sectors will be required to achieve even larger
reductions in order to compensate those sectors. The appropriate
reduction goal for research is a political question that needs to be
addressed by a wide-ranging democratic decision process that that
weighs the GHG emissions from research activities against its societal
benefits. Evidently, this discussion should not be restricted to the
research community alone.\\

\noindent The distribution of our GHG emissions suggests the contours
of a reduction plan. The dominant contribution of astronomical
research infrastructures clearly sets the long-term goal of a
decarbonization strategy, in an effort going well beyond IRAP and
challenging our culture of research. Locally, the quantitative carbon
accounting presented above suggests some non-neligigible emission
reductions that could be achieved in the short term via measures that
would not be technically difficult to implement. As outlined in Ref. \cite{ademe:2014},
a successful emissions reduction plan should combine both aspects:
initial steps that quickly achieve visible emission reductions to set
the organization in motion, and a longer term strategy that attacks
the bulk of the emissions footprint.\\

\noindent A local reduction plan should be carefully co-constructed by
the institute's direction and staff. We thus refrain from making
strong recommendations or listing priorities here. Instead, we
illustrate the magnitude of the emissions reduction that could be
expected from some actions that could be implemented immediately with
limited impact on our daily research activities. Table
\ref{tab:reduction} summarizes these for a series of possible
reduction scenarios:
\begin{enumerate}
\item Professional travel: limiting the number of trips per year and
  per individual in and out of Europe and/or imposing train travel for
  all domestic travel; the estimated reduction does not include the
  cost of hotel accommodation.
  \item Commuting: changing our transport habits by shifting to less
  carbon-intensive means for short distances, like walking or cycling,
  and reducing the remaining mileage done by car by diverting a
  fraction of it towards carpooling and public
  transportation (which could be even more effective if an increased
  fraction of personal cars become electric or hybrid as anticipated
  in France in national reduction strategies), or by extending the
  practice of remote working (accounting for rebound emissions under
  the hypothesis of no adjustments in the lab space usage).
\item Meals: reducing the meat content in our food by progressive
  shifting to flexitarian, flexitarian and vegetarian, or vegetarian
  only meals.
\item Computer equipment: using laptops and other personal computers
  or workstations over longer times, shifting from an average value of
  about 2 years in 2019 to 4 and 6 years; this does not apply to
  servers or tablets or standalone screens.
\item Heating and electricity: reducing our consumption following the
  objectives set for the buildings sector in the ``Strat\'egie
  Nationale Bas Carbone (SNBC)''; the exact scenario for achieving
  this is not specified but would certainly include a better thermal
  insulation and some optimization of the power distribution.
\end{enumerate}

\noindent The solutions considered in Table \ref{tab:reduction}
outline a possible path towards a 20-30\% reduction of our footprint
over the restricted perimeter. Unsurprisingly, the major step would
come from the evolution of our traveling habits. Such changes,
however, are not so drastic since emissions from professional travel
are very unevenly distributed at IRAP, with 20\% (50\%) of the total
amount being due to 12 (45) people only. Such a concentration should
in principle make it easier to efficiently tackle the problem. On top
of this, an ensemble of smaller measures can provide a total reduction
of about 5\%, by acting on the lifetime of computer equipment or
adopting different practices for commuting and meals, while an
ambitious renovation plan for our buildings would lead to a 4\%
benefit if we achieve the national objectives of the SNBC.\\
 
\noindent Even when we ignore the contribution of astronomical research
infrastructures and focus on the emissions associated with IRAP's
restricted perimeter, these measures would not achieve the target of a
50\% reduction in GHG emissions by 2030. This confirms that meeting
the ambitious goal recommended by IPCC for the coming decades will
necessarily imply deep changes in our research culture. Over the
restricted perimeter most closely controlled by IRAP, i.e. the
\gestot\tco, achieving such reduction target will require reducing the
emissions associated with the purchase of goods and services that
supply IRAP's instrument development projects. This presents an
unavoidable challenge to our current relation to that facet of our
activity. We cannot act on the offer of goods and services, but we can
modify our demand by: (i) introducing clauses in purchase rules to
disfavor/exclude suppliers not meeting certain environmental
standards; (ii) restructuring our activities so as to favor less
carbon-intensive purchases. These solutions go beyond IRAP and involve a
progressive shift of standards that should be promoted and supported
at the institutional level.\\

\noindent In France, the SNBC describes strategies to achieve a
reduction of GHG emissions from the industrial sector by 35\% by 2030,
from a reference level in 2015. While this does not meet the 40-60\%
goal set by IPCC, it would be step in the right direction, and would
contribute to reducing the emissions associated with the purchase of
goods and services at IRAP. The impact of the SNBC at IRAP, however,
depends on the fraction of purchases coming from French suppliers. We
note, moreover, that SNBC has failed to meet its objectives over
2015-2018\footnote{https://www.ecologie.gouv.fr/suivi-strategie-nationale-bas-carbone},
with an overall decrease of emissions by 1\% per year whereas more
than 2\% were predicted, which mechanically raises the effort to be
made in subsequent years. It is increasingly clear that we cannot rely
on the decarbonization of our suppliers and partners from other
economic sectors to achieve a 50\% reduction in GHG emissions at IRAP by
2050.\\

\noindent Formally at least, the predicament at IRAP is simple and is
in no way specific to scientific research. As illustrated above, our
carbon footprint is the product of our activity data and the carbon
intensity of those activities. Reducing IRAP's carbon footprint can be
achieved in three different ways: lowering the carbon intensiveness of
our existing activities, reducing the pace of our activities, and
changing our activities towards low-carbon alternatives. We believe we
should make use of all possible levers because: (i) the magnitude of
the reduction to be achieved by society within a decade requires us to
act on all possible aspects of the problem; (ii) reducing and/or
shifting the activity is directly under our control; and (iii)
modifying our activities can have quick and direct effects, as opposed
to the uncertain decarbonization trajectories of suppliers and partner
organizations. These alternatives become especially acute when we
consider the full scope of IRAP's GHG emissions,
rather than the restricted perimeter. \\

\noindent We emphasize that the carbon intensity of the construction
and operation of large research infrastructures is already relatively
low : 140-250\tcopermeuro (see Sect.~\ref{res:infra} and
\cite{Knodlseder:2022} for more detail). This is at the low end of
sector-based emission factors publised by ADEME for a broad range of
activities, from $\sim$100\tcopermeuro for tertiary activities with
little material or technical input, to $\sim$2000\tcopermeuro for
heavy industries \cite{ademe-ratios}. In other words, instrument
development and the operation of telescopes already has a relatively
low carbon intensity. Our massive footprint instead comes from the
large and growing number of facilities we have at our disposal.\\

\noindent Our recommendation for a community-based reduction strategy
would therefore be to divert a growing fraction of our budgets to fund
the decarbonization of existing operational infrastructures, to pursue
research and development of low-carbon technologies on which future
projects will be based, and to reduce the cadence and scale of the
deployment of new research infrastructures. The latter point cannot be
left out of the equation, otherwise any benefit in decarbonizing
existing facilities will promptly be annihilated by an increase in the
number of facilities. An effective, long-term emissions reduction plan
for IRAP (and astronomy research more widely) requires difficult
decisions that must be made today, including for projects that are
already under study. The timescales involved in the development of
astronomical research infrastructures lock in our emissions for the
next decades, and the problem will only be exacerbated if we continue
to postpone the implementation of a far-reaching emissions reduction
strategy.\\

\newpage
% Acknowledgements
\section{Acknowledgements}
\label{acknow}

\noindent We acknowledge the many colleagues who participated in this
effort. We are grateful to the director of the institute, Philippe
Louarn, for supporting this initiative from the very start. We warmly
thank our colleagues from the administration for helping us to collect
the necessary data (by alphabetical order): Marie-Pierre Arberet,
Teddy Ba\"i, St\'ephanie Baylac, Pauline B\'enard, S\'elim Benguesmia,
Jean-Fran\c cois Botte, Sylvie Cieutat, Marjorie Cloup, Christelle
Feugeade, Josette Garcia, Jean-Louis Lefort, Dani\`ele Millet. We also acknowledge financial support from IRAP and CNRS for the training to the ``Bilan Carbone'' methodology.

\bibliographystyle{abbrv}
\bibliography{bilan-ges_irap}

\end{document}